\documentclass[12pt]{article}

\usepackage{amsfonts}
\usepackage{amsmath}
\usepackage{amsthm}
\usepackage[driverfallback=dvipdfm]{hyperref}
\usepackage{cite}
\usepackage{epsfig}
\usepackage{latexsym}
\usepackage{paralist}
\usepackage{fancyhdr}
\usepackage{graphicx}
\numberwithin{equation}{section}
\usepackage[vcentermath]{youngtab}
\usepackage{young}
\usepackage{ytableau}
\usepackage{etex}
\usepackage{braket}
\usepackage{float}
\usepackage{tikz}
\usetikzlibrary{calc}
\usetikzlibrary{topaths}
\usetikzlibrary{decorations}
\usetikzlibrary{decorations.pathmorphing}
\usepackage{autobreak}
\newcommand{\bld}{\boldsymbol}

\usepackage{pict2e}   
\usepackage{animate}  

\allowdisplaybreaks[1]

\def\be{\begin{equation}}
\def\ee{\end{equation}}
\def\bea{\begin{eqnarray}}
\def\eea{\end{eqnarray}}

\renewcommand{\thefootnote}{\fnsymbol{footnote}}

\begin{document}

\hfuzz=100pt
\title{{\Large \bf{Low-Energy Dynamics of \\ 3d $\mathcal{N}=2$ $G_2$ Supersymmetric Gauge Theory   }}}
\date{}
\author{ Keita Nii$^a$\footnote{nii@itp.unibe.ch}
, 
Yuta Sekiguchi$^a$\footnote{yuta@itp.unibe.ch}
}
\date{\today}

\maketitle

\thispagestyle{fancy}
\cfoot{}
\renewcommand{\headrulewidth}{0.0pt}

\vspace*{-1cm}
\begin{center}
$^{a}${{\it Albert Einstein Center for Fundamental Physics }}
\\{{\it Institute for Theoretical Physics
}}
\\ {{\it University of Bern}}  
\\{{\it  Sidlerstrasse 5, CH-3012 Bern, Switzerland}}

\end{center}

\begin{abstract}
We study a three-dimensional $\mathcal{N}=2$ supersymmetric $G_2$ gauge theory with and without fundamental matters. We find that a classical Coulomb branch of the moduli space of vacua is partly lifted by monopole-instantons and the quantum Coulomb moduli space would be described by a complex one-dimensional space.
Depending on the number of the matters in a fundamental representation, the low-energy dynamics of the theory shows various phases like s-confinement or quantum merging of the Coulomb and the Higgs branches. We also investigate superconformal indices as an independent check of our analysis.
\end{abstract}

\renewcommand{\thefootnote}{\arabic{footnote}}
\setcounter{footnote}{0}

\newpage
\tableofcontents 
\clearpage

\section{Introduction}

Supersymmetry allows us to understand the dynamics of strongly-coupled gauge theories. Especially, holomorphy and non-renormalization theorems put quite strong constraints on SUSY dynamics and we can extract non-perturbative effects exactly \cite{Seiberg:1994bz, Seiberg:1994pq, Seiberg:1997vw}. Since a supersymmetry is a symmetry between bosons and fermions, SUSY theories typically contain scalar fields. The potential of these scalar modes  often possesses flat directions and this subspace is called a moduli space of vacua. Then we can introduce vacuum expectation values to these massless modes. This fact very simplifies the SUSY dynamics because by taking the large vevs we can analyze the theory at a semi-classical limit and extrapolate the strongly-coupled region. So it is important to understand the structure of the moduli space of vacua. In addition to this brilliant property, the supersymmetry is regarded as one of the most prevalent theories for Beyond the Standard-Model. Therefore it would be interesting and important to investigate possible SUSY theories and their dynamics including such as SUSY breaking scenarios.

The aim of this paper is to study non-perturbative aspects of $G_2$ gauge theories.
$G_2$ gauge theories have been studied continuously for the last 15 years. For non-supersymmetric $G_2$ cases, see \cite{Holland:2002vk, Holland:2003jy, Pepe:2006er, Cossu:2007dk, Wellegehausen:2010ai, Bruno:2014rxa, Poppitz:2012nz}. Since the $G_2$ group has a trivial center, it is interesting to investigate its confinement phenomena. A confinement phase is usually related to the center symmetry. Wilson loops become well-behaved order parameters in pure gauge theories with a nontrivial center. However, in $G_2$ pure YM theories, Wilson loops in any representations are not well-defined order parameters since the Wilson loops are always screened by gluons.  Also in a $G_2$ QCD with and without higgs fields, 
the confinement and Higgs phases are continuously connected and we have no order parameter to distinguish them. These situations are similar to an ordinary $SU(3)$ QCD and SQCD with fundamental (s)quarks. Furthermore $G_2$ has an $SU(3)$ group as a maximal subgroup. By breaking $G_2$ to $SU(3)$ via the higgs field, the adjoint representation is decomposed into $\mathbf{8}+\mathbf{3}+\bar{\mathbf{3}}$. Therefore we can connect the $G_2$ dynamics to the ordinary vector-like QCD.

In 4d, a supersymmetric version of the $G_2$ theory was also well investigated. The $\mathcal{N}=2$ supersymmetric $G_2$ gauge theory (called a Seiberg-Witten theory) was studied in \cite{Alishahiha:1995wm, Landsteiner:1996ut}, where the Seiberg-Witten curve and their singularities are studied. A 4d $\mathcal{N}=1$ $G_2$ gauge theory was investigated in \cite{Pesando:1995bq, Giddings:1995ns, Smilga:1998em, Davies:2000nw, Alishahiha:2003hj, Saito:2007ah, Bourget:2015upj}, where it was found that the $G_2$ dynamics is similar to the 4d $\mathcal{N}=1$ $SU(N)$ SQCD. Depending on the number of fundamental matters, there are various phases. For $N_f=0$, there are discrete SUSY vacua while for $N_f=1,\cdots,3$ there are no stable SUSY vacua. For $N_f=4$, we observe the quantum-deformed moduli space and for $N_f=5$ the theory is s-confined. For $N_f \ge 6$ we have a Seiberg dual description. Not limited to the development of the SUSY $G_2$ gauge theories, in 4d, other SUSY exceptional gauge theories were also well-studied \cite{Ramond:1996ku, Distler:1996ub, Karch:1997jp, Cho:1997am, Pouliot:2001iw, Csaki:1997aw}\cite{Danielsson:1995zi, Abolhasani:1996np}.


Recently dynamics of the 3d $\mathcal{N}=2$ SUSY gauge theories has been better  understood.
One of the most prominent developments is a localization calculation of SUSY-preserving quantities, such as partition functions, superconformal indices, supersymmetric Wilson loops and so on. 
Using the exact results of these quantities, we can test various conjectures such as AdS/CFT correspondence, Seiberg dualities, mirror symmetry, etc., and one can even discover unknown dualities.
The other progress is a derivation of the 3d (Seiberg) dualities from 4d dualities \cite{Aharony:2013dha, Aharony:2013kma}, where the discussion put the emphasis on the significance of the twisted instantons which appear when putting a 4d theory on a circle. 
In 3d, there are some new properties which are absent in 4d. For instance, we can introduce Chern-Simons terms and real masses by background gauging the global symmetries. The 3d vector superfields supply scalar fields whose potential is (classically) flat and this would be a new modulus absent in 4d. The 3d gauge coupling is relevant even for $U(1)$ gauge theories and we can expect non-trivial dynamics of the $U(1)$. In 3d, there are various dualities known, including the dualities with and without Chern-Simons terms. By connecting the 3d and 4d dualities and their dynamics a la \cite{Aharony:2013dha, Aharony:2013kma}, we can obtain a clear and unified understanding of the SUSY gauge theories in diverse dimensions. While these developments are mostly achieved for the theories with classical Lie groups, the understanding of the 3d $\mathcal{N}=2$ exceptional gauge theories has been less established.

In view of the above situations, we will give a detailed analysis on a 3d $\mathcal{N}=2$ supersymmetric $G_2$ gauge theory with and without fundamental matters. This is a first step for understanding the 3d $\mathcal{N}=2$ exceptional gauge theories and would be a good representative example since $G_2$ is the most simplest exceptional group as far as we have known. 
We first classically analyze the moduli space of vacua of the $G_2$ theory and then turn on quantum effects mostly arising from monopole-instantons. Those non-perturbative effects are exactly determined by virtue of holomorphy and various consistency checks via deformations. Especially the connection with 3d $\mathcal{N}=2$ $SU(3)$ SQCD would be a good test of our study.
We find non-perturbative superpotentials consistent with all the symmetries as in 4d $G_2$ cases \cite{Pesando:1995bq, Giddings:1995ns}. We will find that for $N_f \le 2$ the theory has no stable SUSY vacua, for $N_f=3$ the classical moduli spaces are quantum-mechanically merged and for $N_f=4$ we will encounter so-called an ``s-confinement" phase, where $N_f$ is a number of fundamental matters.
We also study superconformal indices of the theory and discuss spectrum of the low-lying operators (or states). The superconformal indices also give us a non-perturbative check of our treatment. Especially, this will confirm a structure of the quantum Coulomb branch which would be drastically different from the classical picture.

The rest of this paper is organized as follows. In Section 2, we briefly review the dynamics of 4d $\mathcal{N}=1$ supersymmetric $G_2$ gauge theories and define some notations. More complete notations we will use are given in Appendix.
In Section 3, we discuss the Coulomb branch of the moduli space of vacua in the 3d $\mathcal{N}=2$ SUSY gauge theory especially focusing on the $G_2$ case.
In Section 4 and 5, we investigate quantum aspects of the 3d $\mathcal{N}=2$ $G_2$ gauge theory with and without fundamental matters.
In Section 6, the connection between the 3d and 4d $G_2$ gauge theories is investigated.
In Section 7, we compute the superconformal indices for 3d $\mathcal{N}=2$ $G_2$ gauge theories and this would be a non-trivial check of our analysis. 
In Section 8, we will summarize our findings and discuss possible future directions.

\section{Review of 4d $\mathcal{N}=1$ $G_2$ SQCD}
We will briefly review the results of the 4d $\mathcal{N}=1$ supersymmetric $G_2$ gauge theory with $N_f$ fundamental matters \cite{Pesando:1995bq,Giddings:1995ns}. Since a coefficient of the one-loop beta-function is given by $b= 12-N_f$, the theory is asymptotically free for $N_f < 12$. The matter contents and their quantum numbers are summarized in Table \ref{4dG2}, where $\eta_{N_f}$ is a dynamical scale of a $G_2$ gauge coupling, $Q$ is a chiral superfield in a fundamental representation and $\lambda$ is a gaugino in a vector superfield. We listed the generic R-charge in Table \ref{4dG2} and of course, the infrared $U(1)_R$ charge is different from this value.
\begin{table}[H]\caption{Quantum numbers of the 4d $\mathcal{N}=1$ $G_2$ gauge theory} 
\begin{center}
  \begin{tabular}{|c|c||c|c|c| } \hline
   &$G_2$&$SU(N_f)$&$U(1)$&$U(1)_R$ \\ \hline 
   $Q$&${\tiny \yng(1)} \, (\mathbf{7})$&${\tiny \yng(1)}$&1&$R$  \\
 $\lambda$&$\mathrm{adj.} \, (\mathbf{14})$&1&0&1 \\ \hline  
 $\eta_{N_f}=\Lambda^b$&1&1&$2N_f$&$2N_f(R-1) +8$ \\ \hline
  \end{tabular}
  \end{center}\label{4dG2}
\end{table}
Notice that we are listing the anomalous $U(1)$ and $U(1)_R$ symmetries, therefore the linear combination of these $U(1)$'s becomes a genuine $U(1)_R$ symmetry and the other global $U(1)$ is spurious in 4d due to a chiral anomaly. Since we are interested in a 3d theory, we will use this charge assignment for the rest of our paper.
Since a $G_2$ group has a (real) fundamental representation with dimension $7$ and there are three independent invariant tensors; $\delta_{ab}, ~f_{abc},~\tilde{f}_{abcd}$, we can construct the following gauge invariant operators from the chiral superfields (Table \ref{4dG2GI}). $B_{ijk}$ is possible for $N_f \ge 3$ and $F_{ijkl}$ is for $N_f \ge 4$.

\begin{table}[H]\caption{Gauge invariants of the 4d $\mathcal{N}=1$ $G_2$ gauge theory} 
\begin{center}
  \begin{tabular}{|c||c|c|c| } \hline
   &$SU(N_f)$&$U(1)$&$U(1)_R$ \\ \hline 
   $M:= QQ$&${\tiny \yng(2)}$&2&$2R$  \\
 $B:=Q^3$&${\tiny \yng(1,1,1)}$&3&$3R$ \\[8pt]
 $F:=Q^4$&${\tiny \yng(1,1,1,1)}$&$4$&$4R$ \\[8pt] \hline
  \end{tabular}
  \end{center}\label{4dG2GI}
\end{table}

In the following we briefly sketch the quantum dynamics depending on the number of fundamentals. For more detailed analyses, see \cite{Pesando:1995bq, Giddings:1995ns, Smilga:1998em, Pouliot:1995zc, Davies:2000nw}.

\subsubsection*{$N_{f} =  0$: discrete SUSY vacua}
Let us first consider the pure $G_2$ Super Yang-Mills (SYM) theory. The theory supports four discrete supersymmetric vacua \cite{Pesando:1995bq, Smilga:1998em, Davies:2000nw, Holland:2003jy} (see also \cite{Witten:1982df, Morozov:1987hy, Witten:1997bs}). The superpotential is given by 
\begin{align}
W \sim \pm \eta^{1/4}, ~\pm i  \eta^{1/4},
\end{align}
where we omitted numerical factors for simplicity and only kept the fourth root of unity.

\subsubsection*{$N_{f} = 1,2$: Gaugino conedensation}
Next we move on to the $G_2$ SQCD with one or two fundamentals.
The superpotential is dynamically generated by gaugino condensation. For $N_{f} =1,2$, the superpotential consistent with all the symmetries takes
\begin{equation}
W = \left(   \frac{\eta_{N_f}}{ \mathrm{det} \, M}\right)^{\frac{1}{4-N_f}},
\end{equation}
and there is no stable SUSY vacua.

\subsubsection*{$N_{f} = 3$: Instanton generated superpotential}
In this case, the dynamically generated superpotential is again allowed although we have the cubic baryonic branch labeled by $B \equiv  \frac{1}{3!}f_{abc}Q^{a}Q^{b}Q^{c}$. In the case of three flavors, via the generic vacuum expectation value on the Higgs branch, the $G_2$ gauge group is completely broken and the semi-classical calculation of the instanton is justified. As a result, we obtain
\begin{equation}
W = \frac{\eta_{3}}{\mathrm{det}M_{ij} - B^{2}},
\end{equation}
which again has no stable SUSY vacua.

\subsubsection*{$N_{f} = 4$: quantum deformed moduli space}
For $N_f=4$, the quartic baryon $F \equiv \frac{1}{4!}\tilde{f}_{abcd}Q^a Q^b Q^c Q^d$ can be constructed and we classically have some constraints between the mesonic, cubic-baryonic and quartic-baryonic operators. The classical moduli space is quantum mechanically corrected and the origin of the moduli space of vacua is lifted;
\begin{align}
\mathrm{det}M - F^{2} - B^{i}M_{ij}B^{j} = \eta_{4}\,.
\end{align}
The origin of the moduli space of vacua is lifted and some of the symmetries are inevitably broken in this phase.

\subsubsection*{$N_{f} = 5$: s-confinement}
For $N_f=5$, the classical moduli space, including the classical relation between the gauge invariant chiral superfields, is not modified. Especially the origin of the moduli space remains as the quantum moduli space. Therefore it is called s-confinement where we have the confining phase without symmetry breaking.
The symmetry, holomorphy and mass-deformation arguments lead to the superpotential
\begin{equation}
W = \frac{1}{\eta_{5}} \left(-\mathrm{det} M + \frac{1}{2}B^{ik}B^{jl}M_{ij}M_{kl} + F^{i}M_{ij}F^{j} + \frac{1}{4}   \epsilon_{ijklm} F^{i}B^{jk}B^{lm}\right).
\end{equation}
The classical constraints are represented via the equations of motion for the above superpotential.

\subsubsection*{$N_{f} \geq 6$: Seiberg duality}
For $N_{f} \geq 6$, we expect a non-abelian Coulomb phase and the low-energy dynamics is described by the Seiberg magnetic dual \cite{Pouliot:1995zc} with an $SU(N_f -3)$ gauge group with a superpotential
\begin{align}
W= M \bar{q} \bar{q} s +\bar{q}_0 \bar{q}_0 s +\det \, s.
\end{align}
The matter content includes the anti-fundamentals, a symmetric matter and a symmetric meson which is a gauge singlet. The quantum numbers for those matters and for the dual dynamical scale $\tilde{\eta}$ of the $SU(N_f -3) $ gauge group are listed in Table \ref{4dG2dual}.

\begin{table}[H]\caption{A magnetic dual of the 4d $\mathcal{N}=1$ $G_2$ gauge theory} 
\begin{center}
  \begin{tabular}{|c|c||c|c|c| } \hline
   &$SU(N_f-3)$&$SU(N_f)$&$U(1)$&$U(1)_R$ \\ \hline 
   $\bar{q}$&$\bar{{\tiny \yng(1)}} $&${\tiny \bar{\yng(1)}}$&$-1$&$1-R-\frac{1}{N_f-3}$  \\
 $\bar{q}_0$ & $\bar{{\tiny \yng(1)}}$&1&0& $1-\frac{1}{N_f-3}$ \\
$s$ &$\tiny \yng(2)$&1&0&$\frac{2}{N_f-2}$ \\
$M$&$1$&$\tiny \yng(2)$&2&$2R$  \\ \hline
 $\tilde{\eta}_{N_f}=\Lambda^{\tilde{b}}$&1&1&$-N_f$&$-N_f (R-1) -4$ \\ \hline
  \end{tabular}
  \end{center}\label{4dG2dual}
\end{table}
This dual was first found by using the Seiberg duality of the 4d $\mathcal{N}=1$ $Spin(7)$ gauge theory with spinorial matters in \cite{Pouliot:1995zc}. By giving a vacuum expectation value to the spinorial representation, the $G_2$ Seiberg dual is obtained. Notice again that the $U(1)$ global symmetry is spurious and then under the matching of the baryonic operators
\begin{align}
B &:= Q^3  \leftrightarrow \bar{q}^{N_f- 3} \\
F & := Q^4  \leftrightarrow \bar{q}^{N_f- 4} \bar{q}_0,
\end{align}
this $U(1)$ is not acting properly.

\section{Coulomb branch and Monopole operators}
In this section we explain how we define the Coulomb branch operators and calculate their global charges  (quantum-mechanically we have to construct so-called monopole (creating) operators.). For the monopole in the $G_2$ case and the quantization of the magnetic charges, please see \cite{Weinberg:2006rq, Shnir:2015hha, Matsudo:2016cui} and \cite{Lee:1997vp, Lee:1998vu,  Kapustin:2005py}. 

The moduli space of vacua in 3d $\mathcal{N}=2$ SUSY gauge theories is described by two regions, Higgs and Coulomb branches, where chiral and vector superfields take non-zero vacuum expectation values respectively. Of course, depending on the representation of the chiral superfields and the breaking pattern of the gauge group, the Higgs branch might be called a Coulomb or confinement phase. The Higgs branch is parametrized by the gauge invariant composites of the chiral superfields with some constraints between them. This is the same as the 4d case, so we have three composites $M_{ij}, B_{ijk}~(\mbox{for}~N_f \ge 3)$ and $F_{ijkl}~(\mbox{for}~N_f \ge 4)$ for the $G_2$ case. For $N_f \ge 4$, at a generic point of the Higgs branch, $G_2$ can be completely higgsed.

Let us consider the classical Coulomb branch of the 3d $\mathcal{N} = 2$ supersymmetric $G_{2}$ gauge theory. We need $\mathrm{rank}\,(G_2)=2$ coordinates to describe it. At a generic point of the Coulomb brach, $G_2$ is broken to $U(1) \times U(1)$. For each $U(1)$ factor, a corresponding $U(1)$ vector superfield yields complex one-dimensional Coulomb branch which consists of a real scalar in a vector superfield and a dual photon. The dual photon is Hodge-dual to a gauge field and then it is compact.
We would like to parametrize these Coulomb branches in the language of the UV theory.
A set of operators to describe the Coulomb branch is semi-classically given by  
\begin{align}
V_{\alpha} \simeq \exp \, \left[\, \mathrm{Tr}\,  \left( \phi\, \bld{\alpha}^{\vee} \cdot  \bld{H} \right)\, \right]
\end{align}
where $\alpha^{\vee}$ denotes a dual root defined as 
\begin{align}
\bld{\alpha}^{\vee} =\frac{2\bld{\alpha}}{\left<\bld{\alpha}, \bld{\alpha}\right>}
\end{align}
and $\phi$ is an adjoint scalar in a vector superfield and is valued in the Cartan subalgebra:
\begin{align}
\bld{H} &= (H_{1}, H_{2}),~~ \bld{\phi}= \left(\phi_1, \frac{\phi_2}{\sqrt{3}} \right)\,,\\
\phi &= \bld{\phi} \cdot \bld{H} =\phi_{1}H_{1}+\frac{\phi_{2}}{\sqrt{3}}H_{2}
\end{align}
By using the Weyl symmetry we can choose the following chamber with
\begin{align}
\phi_{1} \geq \phi_{2} \geq 0\,.
\end{align}
In the definition of the Coulomb branch operator we are omitting the gauge coupling dependence for simplicity. Since the superfield completion is manifest, we are not specifying the difference between the scalar fields and chiral superfields. Furthermore, rigorously speaking, we have to dualize the gauge field  to a dual photon and include this into the above scalar field to make one complex field. But in this paper we omit this for simplicity and the dependence of the dual photon can be easily restored.

Depending on how to give a vev to each $\phi_{i}$, there are in principle various regions in the Coulomb branch and also there are many corresponding monopole operators. For each positive root (more correctly for each positive dual root), we obtain
\begin{align}
V_{\alpha} &\simeq \exp(2 \phi_2), &V_{\beta} &\simeq \exp(\phi_1-\phi_2), &V_{\alpha +\beta} &\simeq\exp(3 \phi_1-\phi_2) \nonumber \\
V_{2\alpha+\beta} &\simeq\exp(3 \phi_1+\phi_2), &V_{3 \alpha+\beta} &\simeq \exp(\phi_1+\phi_2),  &V_{3\alpha+2\beta} &\simeq \exp(2 \phi_1) =:Z
\end{align}
where $V_{3\alpha+2\beta}= \exp(2 \phi_1)$ will be of special importance, so we labeled it as $Z$ for later convenience.

\subsection{Callias index theorem and zero-modes}
Fermion zero-modes for each Coulomb branch operator can be counted by using Callias' index theorem \cite{Callias:1977kg, Weinberg:1979zt, deBoer:1997kr}. The index theorem states that the number of zero-modes for fermions in some representations of the gauge group is given by
\begin{align}
N_{\alpha}= \frac{1}{2} \sum_{w \in \mathrm{all\,the\,weights}} \mathrm{sign} (w(\phi) ) \, w(g), 
\end{align}
where $g= \bld{\alpha}^\vee \cdot  \bld{H}  $ represents the magnetic charge of the monopole we are considering and $\phi$ is the coordinates of the Coulomb branch. The summation is taken over all the weights in a representation.

For two roots $\bld{\alpha}$ and $\bld{\beta}$, for example, using 
\begin{equation}
\bld{\alpha}^{\vee} = (0,2\sqrt{3})\,,\quad \bld{\beta}^{\vee} = (1, -\sqrt{3})\,,
\end{equation}
the zero-modes for adjoint and fundamental fermions are computed as
\begin{align}
N^{\mathrm{adj.}}_{\alpha} &= \frac{1}{2}\biggl[ -6\mathrm{sign}(\phi_{1}-\phi_{2}) - 2\mathrm{sign}(3\phi_{1}-\phi_{2}) +4\mathrm{sign}(\phi_{2})    \nonumber \\
&\qquad \qquad + 6\mathrm{sign}(\phi_{1}+\phi_{2})+2\mathrm{sign}(3\phi_{1}+\phi_{2}) \biggr] =2\nonumber \\
N^{\mathrm{fund.}}_{\alpha} &=\frac{1}{2}\biggl[-2\mathrm{sign}(3\phi_{1}-\phi_{2})+4\mathrm{sign}(\phi_{2}) + 2\mathrm{sign}(3\phi_{1}+\phi_{2})\biggr]=2\,,
\end{align}
\begin{align}
N_{\beta}^{\mathrm{adj.}} &= \frac{1}{2}\biggl[2\mathrm{sign}\phi_{1} + 4\mathrm{sign}(\phi_{1}-\phi_{2})+ 2\mathrm{sign}(3\phi_{1}-\phi_{2}) - 2\mathrm{sign}(\phi_{2}) - 2\mathrm{sign}(\phi_{1}+\phi_{2})\biggr]=2\nonumber\\
N_{\beta}^{\mathrm{fund.}} &=\frac{1}{2}\biggl[2\mathrm{sign}(3\phi_{1}-\phi_{2})-2 \mathrm{sign}\phi_{2}\biggr]=0\,,
\end{align}
where the sign function is evaluated under the Weyl chamber. In the presence of the monopole vertex, the naive global symmetries are broken because the vertex contains the fermions corresponding to the zero-modes above. In order to recover the global symmetries, we have to transform the monopole operator in a opposite way to the fermions under the global symmetry \cite{Affleck:1982as}. Since the Coulomb branch operators are made from the vector superfields, they are originally neutral.  However, on the monopole background, they are non-trivially charged. Consequently, the operator $V_{\alpha}$ possesses a $U(1)_{R}$-charge
\begin{align}
R[V_\alpha]=-2 \cdot R[\lambda] - 2\cdot N_{f}\cdot R[\psi_Q] = 2N_{f}(1-R)-2\,,
\end{align}
while for $V_{\beta}$ and $Z = V_{3\alpha+2\beta}$ 
\begin{align}
R[V_{\beta}] &=-2\cdot R[\lambda] =  -2\\
R[Z] &=R[V_\alpha V_\beta^2] = 2N_{f}(1-R) - 6\,.
\end{align}
The other number of zero-modes for each operator is summarized in Table \ref{tb: zeromodes}. It is remarkable to note that the $U(1)_{R}$-charge of the operator $V_{\beta}$ depends neither on $N_{f}$ or R, due to the absence of zero modes in fundamental representations. This implies that the inverse of $V_{\beta}$ will be ubiquitous in the superpotential for any number of flavors. We will discuss the uplift of the $V_{\beta}$-direction in the next section.

\begin{table}[H]\caption{Zero-modes for Coulomb branch operators and global charges}
\begin{center}
  \begin{tabular}{|c|c|c|c|c|c| } \hline
   & adj. & fund. &U(1) & $U(1)_{R}$ \\ \hline 
   $V_{\alpha}$ &2 & 2 &$-2N_f$& $2N_{f} (1- R) -2$\\
   $V_{\beta}$&2&0 &0&$-2$\\
  $V_{\alpha +\beta} $ &8&2&$-2N_f$&$2N_{f} (1- R)-8$\\
 $ V_{2\alpha+\beta}$ &10&4&$-4N_f$&$4N_{f} (1- R)-10$\\
  $V_{3 \alpha+\beta} $ &4&2&$-2N_f$&$2N_{f} (1- R)-4$ \\
 $Z =V_{3\alpha+2\beta}= V_{\alpha}V^2_{\beta}$& 6& 2 &$-2N_f$& $2N_{f}(1-R) -6$\\\hline 
  \end{tabular}
  \end{center}\label{tb: zeromodes} 
\end{table}

\subsection{Mixed Chern-Simons terms and zero-modes}
Since the number of fermionic zero-modes can be also studied via mixed Chern-Simons terms \cite{Intriligator:2013lca}, we here give an alternative argument of deriving the global charges for the Coulomb branch operators. But one can easily find that this is equivalent to the above calculation.

Let us first calculate the charges of the monopole operator $Z\simeq \exp(2 \phi_1)$. Along the moduli of a non-zero value of $\braket{Z} $, the gauge group is broken as $G_2 \rightarrow SU(2) \times U(1)$. The $Z$ direction corresponds to a dual root of $3\alpha +2 \beta$ which is perpendicular to the root $\alpha$. Therefore the $SU(2)$ with the roots $\alpha, -\alpha$ and a Cartan generator $H_2$ remains unbroken and a $Z$ direction corresponds to the unbroken $U(1)$ related to $H_1$. Under this breaking, the fields are decomposed as
\begin{align}
\mathbf{7} &\rightarrow \mathbf{3}_0+ \mathbf{2}_{1} + \mathbf{2}_{-1} \\
\mathbf{14} &\rightarrow \mathbf{3}_0 +\mathbf{1}_0 +\mathbf{4}_1 +\mathbf{4}_{-1} +\mathbf{1}_2 +\mathbf{1}_{-2} 
\end{align}

In order to calculate the effective Chern-Simons terms, we have to know the sign of the masses of fermions which appear in 1-loop graphs.
The mass terms for the fermions are dictated from
\begin{align}
\phi &= \phi_1 H_1 +\frac{\phi_2}{\sqrt{3}} H_2 \nonumber \\ 
&= \frac{1}{2} \begin{pmatrix}
 0 & 0 & 0 & 0 & 0 & 0 & 0\\
 0 & -\phi_1 -\frac{1}{3} \phi_2 & 0 & 0 & 0 & 0 & 0\\
 0 & 0 & -\frac{2}{3} \phi_2 & 0 & 0 & 0 & 0\\
 0 & 0 & 0 & -\phi_1 +\frac{1}{3} \phi_2 & 0 & 0 & 0\\
 0 & 0 & 0 & 0 & \frac{2}{3} \phi_2 & 0 & 0\\
 0 & 0 & 0 & 0 & 0 & \phi_1 + \frac{1}{3} \phi_2 & 0\\
 0 & 0 & 0 & 0 & 0 & 0 & \phi_1 -\frac{1}{3} \phi_2
\end{pmatrix}.
\end{align}
 Then the mixed Chern-Simons terms are
\begin{align}
k_{eff}^{U(1)_{gauge} U(1)_{global}} &=\frac{1}{2} N_f\left[ \mathrm{sign} \left(\phi_1 + \frac{\phi_2}{3}\right) + \mathrm{sign} \left( \phi_1 - \frac{\phi_2 }{3} \right) \nonumber \right. \\
&\left.  \qquad \qquad \quad - \mathrm{sign}  \left(-\phi_1 +  \frac{\phi_2}{3}  \right) - \mathrm{sign}  \left(-\phi_1 - \frac{\phi_2}{3}  \right) \right] =2N_f 
\end{align}
\begin{align}
&k_{eff}^{U(1)_{gauge} U(1)_{R}} =2N_f (R-1) +\frac{1}{2} \biggl[ 2 \mathrm{sign} (\phi_1) -2\mathrm{sign} (-\phi_1)   \nonumber \\
& \qquad \qquad  \qquad +\mathrm{sign}(\phi_1+ \phi_2) +\mathrm{sign} \left(\phi_1+ \frac{\phi_2}{3} \right)  +\mathrm{sign}\left( \phi_1- \frac{\phi_2}{3} \right)  +\mathrm{sign}  ( \phi_1- \phi_2)  \nonumber \\
&\qquad \qquad  \qquad \left. - \mathrm{sign}(-\phi_1-\phi_2) -\mathrm{sign}  \left(-\phi_1- \frac{\phi_2}{3} \right)  -\mathrm{sign} \left( -\phi_1+\frac{\phi_2}{3} \right) -\mathrm{sign}(-\phi_1+ \phi_2) \right] \nonumber \\[5pt]
&\qquad \qquad  \qquad =2N_f (R-1)  + 6
\end{align}
Notice that the CS term for $U(1)_{gauge}$ is vanishing and this is consistent with the fact that the monopole operator $Z$ is gauge-invariant.

Next we consider the operator $V_\alpha \simeq \exp(2 \phi_2)$. Along this direction, an $SU(2)$ with $3 \alpha +2 \beta, -(3\alpha +2\beta)$ and $H_1$ remains unbroken. The operator $V_\alpha$ corresponds to the monopole-creating operator with a $U(1)$ from $H_2$. Under the breaking $G_2 \rightarrow SU(2) \times U(1)_{H_2}$, the fundamental and adjoint fields are decomposed as 
\begin{align}
\mathbf{7} &\rightarrow \mathbf{2}_{1}+\mathbf{2}_{-1}  +\mathbf{1}_{2}+\mathbf{1}_{0} +\mathbf{1}_{-2}  \\
\mathbf{14} &\rightarrow \mathbf{3}_{0}+\mathbf{1}_0 + \mathbf{2}_{ \pm 3} +\mathbf{2}_{\pm 1} +\mathbf{1}_{\pm 2 }.
\end{align}
By carefully taking into account the mass term for each representation, we find
\begin{align}
&k_{eff}^{U(1)_{gauge} U(1)_{global}} = \frac{1}{2} N_f \left[ \mathrm{sign}\left(\phi_1 + \frac{1}{3} \phi_2  \right)  +\mathrm{sign}\left(-\phi_1 + \frac{1}{3} \phi_2  \right)  \right. \nonumber \\
&\qquad \qquad \qquad \quad \left. -\mathrm{sign}\left(-\phi_1 -\frac{1}{3} \phi_2  \right) -\mathrm{sign}\left(\phi_1 - \frac{1}{3} \phi_2  \right) +2 \mathrm{sign}\left(\frac{2}{3} \phi_2 \right)   -2\mathrm{sign}\left(-\frac{2}{3} \phi_2  \right)  \right] \nonumber \\[5pt]
&\qquad \qquad \qquad \quad =2N_f  
\end{align}
\begin{align}
&k_{eff}^{U(1)_{gauge} U(1)_{R-charge}} =2N_f (R-1) +\frac{1}{2} \biggl[  3 \mathrm{sign}(\phi_1+\phi_2) +3\mathrm{sign}(-\phi_1+\phi_2)  \nonumber \\  
& \qquad \qquad \qquad \quad \left. -3\mathrm{sign}(-\phi_1-\phi_2)   -3\mathrm{sign}(\phi_1-\phi_2)+ \mathrm{sign} \left(\phi_1+{\phi_2 \over 3}  \right)    + \mathrm{sign} \left(-\phi_1+ { \phi_2 \over 3 } \right)  \right. \nonumber \\
& \qquad \qquad \qquad \quad  \left.  -\mathrm{sign}  \left(\phi_1-{ \phi_2  \over 3}  \right) - \mathrm{sign}  \left(-\phi_1- {  \phi_2 \over 3}  \right)   +2 \mathrm{sign} \left({ \phi_2  \over  3} \right)-2 \mathrm{sign}  \left(-  {\phi_2  \over 3}  \right)            \right] \nonumber \\[8pt]
&\qquad \qquad \qquad ~~~~ \quad =2N_f (R-1) + 2
\end{align}

Finally we study the Coulomb branch with $V_\beta \simeq \exp(\phi_1-\phi_2)$. The symmetry breaking is schematically
\begin{align}
G_2  &\rightarrow SU(2)_{\pm(2 \alpha +\beta),H_1+H_2/\sqrt{3} } \times U(1)_{H_1-\sqrt{3}H_2 }
\end{align}
and the branching rules for the fundamental and adjoint representations are
\begin{align}
\mathbf{7}  & \rightarrow  \mathbf{3}_{0} + \mathbf{2}_1 +\mathbf{2}_{-1} \\
\mathbf{14} & \rightarrow  \mathbf{3}_0 +\mathbf{1}_0 +\mathbf{4}_1 +\mathbf{4}_{-1} +\mathbf{1}_2 +\mathbf{1}_{-2}.
\end{align}
Then we can compute the mixed Chern-Simons terms generated along this direction:
\begin{align}
k_{eff}^{U(1)_{gauge} U(1)_{global}} & =  \frac{1}{2} N_f  \left[ \mathrm{sign} \left(- {2\phi_2 \over 3} \right)+ \mathrm{sign} \left(\phi_1 - {\phi_2 \over 3 } \right)  \right.  \nonumber \\
&\left. \qquad \qquad \qquad \qquad  -\mathrm{sign} \left(-\phi_1 +{\phi_2 \over 3}  \right) - \mathrm{sign} \left( { 2\phi_2 \over 3}  \right)  \right]  =0 \\
k_{eff}^{U(1)_{gauge} U(1)_{R-charge}} &=  \frac{1}{2} \left[ \mathrm{sign} (\phi_1) +\mathrm{sign}\left( \phi_1-{ \phi_2  \over 3} \right) +\mathrm{sign}\left(-{\phi_2\over3} \right) +\mathrm{sign}(-\phi_1-\phi_2) \right.  \nonumber \\
& \quad \qquad  -\mathrm{sign}(\phi_1 +\phi_2) -\mathrm{sign} \left( {\phi_2 \over 3} \right) -\mathrm{sign} \left(-\phi_1 +{\phi_2 \over 3}  \right) -\mathrm{sign}(-\phi_1)   \nonumber  \\
& \qquad \qquad   +2 \mathrm{sign} \left(\phi_1-\phi_2 \right) -2 \mathrm{sign}(-\phi_1+\phi_2) \biggr]  \nonumber \\
&=2
\end{align}
Therefore along the branch $\braket{V_\beta} \simeq \exp ( \phi_1 - \phi_2)$, we have no fundamental fermion zero-mode but two gluino zero-modes should contribute.

\section{3d $\mathcal{N}=2$ $G_2$ pure Yang-Mills}
We will start with quantum considerations of the moduli space of vacua from the pure $G_2$ SYM without fundamental matters. The 3d $\mathcal{N}=2$ pure SYM theories for various gauge groups were studied in \cite{Davies:2000nw} with the connection to the theory in 4d and in $\mathbb{S}^1 \times \mathbb{R}^3$. Since the $G_2$ group has rank 2, the Coulomb branch is classically two-dimensional and these are described by two monopole operators corresponding to the simple roots. We labeled them as $V_\alpha$ and $V_\beta$. The symmetry argument says that the following terms are generated in the superpotential
\begin{align}
W=\frac{3}{V_\alpha} +\frac{1}{V_\beta},
\end{align}
where we are including the relative coefficient related to the length of the roots. This is consistent with \cite{Davies:2000nw}.
These are monopole-generated superpotentials since the monopoles corresponding to the $G_2$ breaking,
\begin{align}
G_2 \rightarrow U(1) \times U(1)
\end{align}
contains two gaugino zero-modes and they can contribute to the superpotential. These terms prevent us from giving the vacuum expectation values to these Coulomb branch directions. If we recall the relation between the monopole operators and the classical Coulomb branch, $V_\beta \simeq \exp(\phi_1- \phi_2) $, the repulsive force is acting between $\phi_1$ and $\phi_2$. So in the Weyl chamber we expect that the $\phi_1$ direction can be turned on while the $\phi_2$ is frozen to zero.  Even if we add the fundamental matters, the direction with $V_\beta$ is still lifted via the monopole superpotential since the fundamental quarks do not have any zero-mode around the $V_\beta$ monopole as we have seen in Section 3. Then it is natural to think that the quantum Coulomb branch is one-dimensional and this would be parametrized by an operator including only the $\phi_{1}$ variable, namely, a $Z\simeq \exp(2 \phi_1)$ operator. The validity of this candidate will be discussed by extending to the inclusion of fundamental flavors in Section 5. In addition, the discussion on the theory on $\mathbb{S}_{1}\times \mathbb{R}_{3}$ will make the operator $Z$ more suitable for the description of the quantum Coulomb branch in Section 6.

\section{3d $\mathcal{N}=2$ $G_2$ SQCD}
Next we introduce chiral superfields in a fundamental representation to the $G_2$ theory discussed above. The matter contents and their representations are sumarized in Table \ref{3dG2SQCD}. Notice that the global $U(1)$ symmetry is not spurious but a genuine symmetry now. Fermion zero-modes from the fundamental matters modify the zero-mode counting for the Coulomb branch operators except for $V_\beta$. Therefore, even for non-zero $N_f$, we can still have the superpotential 
\begin{align}
W={ 1  \over V_\beta}
\end{align}
and this direction would be lifted. It is natural to regard the one-dimensional Coulomb branch of $Z$ as quantum-mechanically massless and as a globally defined monopole-creating operator for non-zero $N_f$. By using $M_{ij}, B_{ijk}, F_{ijkl}$ and $Z$ we can find the following phases and the superpotentials for $N_f \le 4$.

\begin{table}[H]\caption{Quantum numbers of the $G_2$ SQCD with $N_f$ flavors} 
\begin{center}
  \begin{tabular}{|c|c|c|c|c|c|c| } \hline
   &$G_{2}$& $SU(N_{f})$&$U(1)$&$U(1)_{R}$\\ \hline 
   $Q$ & {\tiny \yng(1)} & {\tiny \yng(1)} & 1& $R$\\
  $\lambda$&$\mathrm{adj.}$&1 &0&1\\ \hline
 $M_{ij}=Q_{i}^{a}Q^{a}_{j}$& 1& ${\tiny \yng(2)}$ & $2$ & $2R$\\[5pt]
 $B = Q^{3}$ & 1 & ${\tiny \yng(1,1,1)}$ & 3 & $3R$\\[8pt]
 $F = Q^{4}$ & 1 & ${\tiny \yng(1,1,1,1)}$ & 4 & $4R$\\
 $Z = e^{2\phi_{1}}$ & 1 & 1 & $-2N_{f}$ & $2N_{f}(1-R) -6$\\
   \hline 
  \end{tabular}
  \end{center}\label{3dG2SQCD}
\end{table}

\subsubsection*{$N_f=1$}
When $N_f =1$, the low-energy dynamics is similar to the 4d $\mathcal{N}=1$ $G_2$ gauge theory with $N_f=2$ fundamental matters. The superpotential below is allowed from the symmetry argument.
\begin{align}
W = \left(\frac{1}{M Z} \right)^{\frac{1}{2}}
\end{align}
By differentiating the superpotential, we obtain the runaway potential and there is no stable SUSY vacua.
The consistency can be checked by flowing to the Higgs branch. Along the Higgs branch, the $G_2$ gauge group is broken down to $SU(3)$ which is a maximal subgroup of $G_2$. By introducing the vev $\braket{M}=v$, we again find no SUSY solution. This is consistent with the dynamics of a 3d $\mathcal{N}=2 
$ $SU(3)$ without matters where the monopole corresponding to the breaking $SU(3) \rightarrow U(1) \times U(1)$ creates the runaway potential. 
We can also test this superpotential by introducing a complex mass to the chiral superfield. By integrating the massive modes, we have $W= Z^{-1/3}$ and there is no stable SUSY vacuum.

\subsubsection*{$N_f=2$}
The dynamics of $N_f=2$ is similar to the 4d $\mathcal{N}=1$ $G_2$ theory with $N_f=3$. We again have a runaway-type superpotential. 
\begin{align}
W = \frac{1}{Z\,\mathrm{det}M}
\end{align}
By introducing a vacuum expectation value with $\mathrm{rank} \, \braket{M} =1$, we can flow to a 3d $\mathcal{N}=2$ $SU(3)$ with one flavor. By properly rescaling the Coulomb branch operator $V_{SU(3)} :=2 v Z$ the low-energy dynamics is described by
\begin{align}
W= \frac{1}{ V_{SU(3)} M_{11}},
\end{align}
which explains the dynamics of a 3d $\mathcal{N}=2$ $SU(3)$ with one flavor.
We can also deform the theory by a complex mass. Let us introduce a term $m M_{22}$ and we obtain
\begin{align}
W= \frac{1}{Z \mathrm{det} \, M} +m M_{22} \rightarrow \left( \frac{m}{Z M_{11}}  \right)^{\frac{1}{2}}.
\end{align}
By properly rescaling the monopole operator, we have the superpotential of $N_f=1$.

\subsubsection*{$N_f=3$: quantum deformed moduli space}
The dynamics of a 3d $G_2$ theory with $N_f=3$ is similar to the 4d $\mathcal{N}=1$ $G_2$ gauge theory with 4 fundamental matters and also to the 3d $\mathcal{N}=2$ $SU(N_c)$ gauge theory with $N_c-1$ flavors. We find the following constraint is consistent with all the symmetries.
\begin{align}
Z\,(\mathrm{det}M - B^{2}) = 1
\end{align}
This constraint relates the large values of the Higgs branch to the vicinity of the origin of the Coulomb branch. Thus the Coulomb and Higgs branches are quantum mechanically merged. This constraint also means that some global (and also gauge) symmetries are inevitably broken on the whole moduli space and the origin of the moduli space is not a vacuum.

We can test this phase by considering the Higgs branch. As in the previous case, let us introduce the $\mathrm{rank} \, \braket{M} =1$ vev. In this case the UV theory flows to a 3d $\mathcal{N}=2$ $SU(3)$ gauge theory with 2 flavors at the low-energy limit. The global non-abelian symmetry is enhanced since the $\mathbf{7}$ representations yield $\mathbf{3} +\bar{\mathbf{3}}$. Therefore we have to carefully decompose the gauge invariant operators \cite{Pesando:1995bq}.  The symmetric meson is decomposed to $\hat{M}_{f} ^{~\bar{g}} +\hat{M}_{g}^{~\bar{f}}$ and the cubic baryon becomes $i a (\hat{M}_{2}^{~3}-\hat{M}_3^{~2} )$ where $a$ is a vev of the fundamental squark. By inserting this expression we finally obtain 
\begin{align}
V_{SU(3)} \det \left(  \, \hat{M}_{i}^{~\bar{j}} \right) =1,
\end{align}
where we rescaled the monopole operator and absorbed $a^2$ and the unimportant numerical factor.
This result is precisely the 3d $SU(3)$ result with 2 flavors.

\subsubsection*{$N_f=4$: s-confinement}
Finally, we present the dynamics of $N_f=4$. The phase of the 3d $\mathcal{N}=2$ $G_2$ gauge theory with 4 flavors is similar to a 4d $\mathcal{N}=1$ $G_2$ theory with $5$ flavors where one can see the s-confinement phase. The superpotential consistent with all the symmetries is
\begin{align}
W = Z (- \mathrm{det}M + F^{2} + B^{i}M_{ij}B^{j} ),
\end{align}
where the relative coefficients are chosen as we reproduce the result of $N_f \le 3$ when integrating out the massive flavors by introducing complex masses. The massless excitations are $M_{ij},~B_i,~F$ and the monopole operator $Z$. The interaction between these massless modes are described by the above potential. At the origin of moduli space of vacua (in the present case, the origin belongs to the vacua as different from the $N_f=3$ case.), none of the global symmetries is broken. So this phase is called s-confinement. We can see the consistency by calculating the parity anomaly. For the UV theory, each effective Chern-Simons level between the global $U(1)$ and $U(1)_R$ symmetries is computed as
\begin{align}
k^{UV}_{U(1)_R U(1)_R} &= \frac{1}{2} \left( 7 N_{f}\, \mathrm{sign} \, M_{Q} + 14\, \mathrm{sign} \, M_{\lambda} \right) &\in &\quad \begin{cases}
    \mathbb{Z} +\frac{1}{2}& (\mbox{odd}~N_f) \\
     \mathbb{Z} & (\mbox{even}~N_f)
  \end{cases}  \\
k^{UV}_{U(1) U(1)_R} &=-\frac{7}{2}N_{f}\,\mathrm{sgn}M_{Q} &\in &\quad \begin{cases}
    \mathbb{Z} +\frac{1}{2}& (\mbox{odd}~N_f) \\
     \mathbb{Z} & (\mbox{even}~N_f)
  \end{cases}\\
k^{UV}_{U(1)U(1)} &= \frac{1}{2}N_{f}\,\mathrm{sgn}M_{Q} &\in&\quad \begin{cases}
    \mathbb{Z} +\frac{1}{2}& (\mbox{odd}~N_f) \\
     \mathbb{Z} & (\mbox{even}~N_f)
  \end{cases}.
\end{align}
The similar calculation is performed for the IR description and we find the matching for $N_f=4$:
\begin{align}
k^{IR}_{U(1)_R U(1)_R} &= \frac{1}{2}  \left( 10 \mathrm{sign} \,M_{M} +4\mathrm{sign} \, M_B +\mathrm{sign} \,M_F +\mathrm{sign} \, M_Z  \right) &\in &\quad \mathbb{Z} \\
k^{IR}_{U(1) U(1)_R} &=\frac{1}{2} \left( -20 \mathrm{sign} \, M_M -12\mathrm{sign} \, M_B -4\mathrm{sign} \, M_F -8\mathrm{sign} \, M_Z \right) &\in & \quad \mathbb{Z}\\
k^{IR}_{U(1)U(1)} &= \frac{1}{2} \left(40 \mathrm{sign} \, M_M +36\mathrm{sign} \, M_B +16\mathrm{sign} \, M_F+64\mathrm{sign} \, M_Z\right) &\in & \quad \mathbb{Z}
\end{align}

We can also test this phase by considering the Higgs branch as in \cite{Pesando:1995bq}. By adding the $\mathrm{rank} \, \braket{M}=1$ vev to the theory, we flow to the 3d $\mathcal{N}=2$ $SU(3)$ gauge theory with $N_f=3$ flavors. When flowing to the $SU(3)$ gauge theory, the flavor symmetry is enhanced to $SU(3)_L\times SU(3)_R$. By introducing the vev to the 1st component of the flavor, the gauge invariant composites reduce to
\begin{align}
M_{ij} &=  \hat{M}_i^{~\bar{j}} +\hat{M}_j^{~\bar{i}} \\
B^{1} &= i\sqrt{2} (b-\bar{b}) \\
B^{i} &=  \frac{ia}{2} \epsilon_{1ijk} \left(  \hat{M}_j^{~\bar{k}} -\hat{M}_k^{~\bar{j}} \right), ~~~(i,j,k =2,3,4)
\end{align}
and the superpotential becomes
\begin{align}
W= - 8 a^2 Z (\mathrm{det} \, \hat{M} - b\bar{b} )   = -Y_{SU(3)}   (\mathrm{det} \, \hat{M} - b\bar{b} ). 
\end{align}
where $b$ and $\bar{b}$ are the (anti-)baryonic operators for the $SU(3)$ theory, $\hat{M}$ is a meson with $SU(3)_L \times SU(3)_R$ indices and  $a$ is a vev for $Q$. We rescaled the Coulomb branch as $Y_{SU(3)}:=  8 a^2 Z$. This low-energy superpotential is consistent with the 3d $\mathcal{N}=2$ $SU(3)$ gauge theory with three flavors \cite{Aharony:1997bx}.

\section{$G_2$ SQCD on $\mathbb{S}_1 \times \mathbb{R}_3$}
We can connect the $G_2$ dynamics in 3d and 4d via compactification of the 4d theory on a circle and by taking into account non-perturbative effects from the twisted-monopole \cite{Lee:1997vp, Aharony:1997bx} (it is known also as Kaluza-Klein monopole.). Generally speaking, if we compactify one direction of the space-time, the 4d BPST instanton is called a ``caloron'' (see for example \cite{Nahm:1983sv}.). This caloron configuration can be regarded as the bound state of the magnetic monopoles and the KK-monopole. Since the magnetic monopole has the same number of fundamental fermion zero-modes as the 4d instanton, the KK-monopole only has the adjoint zero-modes in our setup. Therefore the KK-monopole has only two gaugino zero-modes and it can contribute to the superpotential.
In our present case we obtain
\begin{align}
W=\eta Z=\eta V_\alpha V_\beta^2 .
\end{align}
This is consistent with \cite{Davies:2000nw} since the $Z$ direction is related with the lowest co-root.

Let us start with the analysis of the pure SYM on $\mathbb{S}_1 \times \mathbb{R}_3$. We now have two contributions from the magnetic monopoles and the KK-monopole:
\begin{align}
W_{N_f=0} =\frac{3}{V_\alpha} +\frac{1}{V_\beta} +\eta V_\alpha V_\beta^2
\end{align}
Since the Coulomb moduli should be integrated out in a 4d limit, by solving the F-flatness conditions we find four discrete SUSY vacua and the superpotential
\begin{align}
W_{4d~limit}= \pm 2^{3/2} 3^{1/4} \eta^{1/4},~~\pm i  2^{3/2} 3^{1/4} \eta^{1/4},
\end{align}
which explains the gaugino condensation and is consistent with the fact that the 4d $\mathcal{N}=1$ $G_2$ pure SYM has 4 discrete SUSY vacua. The coefficient is a fourth root of unity as it should be \cite{Pesando:1995bq, Smilga:1998em, Davies:2000nw, Holland:2003jy}.

For the theory with fundamental matters, we again obtain the 4d superpotential by integrating out the monopole operator. 
\begin{align}
W_{N_f=1} &=\qquad \frac{1}{(MZ)^{1/2}} +\eta Z &\rightarrow &\quad W_{N_f=1}^{4d}=\quad \left( \frac{\eta}{M} \right)^{\frac{1}{3}} \\
W_{N_f=2} &=\qquad \frac{1}{Z\,\mathrm{det}M} +\eta Z &\rightarrow &\quad W_{N_f=2}^{4d} = ~ \left( \frac{\eta}{\mathrm{det} \, M} \right)^{\frac{1}{2}}\\
W_{N_f=3} &= X ( Z\,(\mathrm{det}M - B^{2}) - 1 )  +\eta Z &\rightarrow &\quad W_{N_f=2}^{4d}=  \,  \frac{\eta}{\mathrm{det}M - B^{2}}  \\
W_{N_f=4} &=  Z (- \mathrm{det}M + F^{2} + B^{i}M_{ij}B^{j} ) +\eta Z &\rightarrow & \quad \mathrm{det}M - F^{2} - B^{i}M_{ij}B^{j} = \eta\quad 
\end{align}
where for $N_f=3$ we introduced a Lagrange multiplier field $X$ to impose the constraint and $X$ is also integrated out from the low-energy spectrum in the 4d limit.

For $N_f =5$, we can write down the ``effective'' superpotential in 3d as
\begin{align}
W_{N_f=5}^{3d} = \left[ Z\,(-\mathrm{det}M + B^{ij}B^{kl}M^{ik}M_{jl} + F^{i}M_{ij}F^{j} + \epsilon_{ijklm}B^{ij}B^{kl}F^{m} ) \right]^{\frac{1}{2}}. 
\end{align}
This superpotential is singular in the origin of the moduli space, which signals that extra massless degrees of freedom should emerge there and we must add new massless modes to this effective description. However we can use this ``effective'' description far away from the origin of the moduli space and one can go back to the 4d theory as follows.
\begin{align}
W_{N_f=5}^{\mathbb{S}^1 \times \mathbf{R}^3 }  &=  W_{N_f=5}^{3d} +\eta Z \nonumber \\
&\rightarrow  \frac{1}{\eta} (-\mathrm{det} \,M + B^{ij}B^{kl}M^{ik}M_{jl} + F^{i}M_{ij}F^{j} + \epsilon_{ijklm}B^{ij}B^{kl}F^{m} )
\end{align}
%

\section{Superconformal Indices}
In this section we calculate the superconformal indices \cite{Bhattacharya:2008bja, Kim:2009wb, Imamura:2011su, Imamura:2011uj, Kapustin:2011jm} (see also \cite{Bashkirov:2011vy, Kapustin:2011vz, Kim:2013cma} and \cite{Spiridonov:2009za}) for 3d $\mathcal{N}=2$ $G_2$ gauge theories and confirm that the previous analysis is correct. Especially we will observe that the quantum Coulomb branch is indeed one-dimensional and described by the monopole operator $Z$. 

The 3d superconformal indices (known as twisted partition functions on $\mathbb{S}_1\times \mathbb{S}_2$) are given by a localization technique \cite{Kapustin:2009kz} and the result is
\begin{align}
I (x,t) &= \sum_{s_1,s_2 }\frac{1}{|\mathrm{Sym}|}  \oint \oint \prod_{i=1,2} \frac{dz_i}{ 2 \pi i z_i} \, Z_{\mathrm{vector}} Z_{\mathrm{chiral}} \nonumber \\
& Z_{\mathrm{vector}} = \prod_{\mathrm{\alpha \in all\,the\,roots}} x^{-|\alpha(s)|} (1- e^{i \alpha(h)} x^{2| \alpha(s)|})  \nonumber \\
& Z_{\mathrm{chiral}} = \prod_{\Phi} \prod_{\mathrm{\rho_\Phi \in all\,the\,weights}} (x^{1-\Delta_\Phi} e^{-i \rho(h) } t^{-1} )^{|\rho(s)|}  \frac{( e^{-i \rho ( h ) } t^{-1} x^{2|\rho(s)| +2-\Delta_\Phi} ; x^2 )_\infty}{( e^{i \rho ( h ) } t x^{2|\rho(s)| +\Delta_\Phi} ; x^2 )_\infty},
 \end{align}
where $(a; x^2)_\infty$ is a q-Pochhammer symbol 
\begin{align}
(a; q)_\infty := \prod_{k=0}^\infty (1-a q^k),
\end{align}
and we introduced the fugacity $t$ only for the global $U(1)$ symmetry for simplicity and it suffices for our purpose. $\Delta_{\Phi}$ is a conformal weight of the chiral superfields. Since we do not know a true value for it, we chose specific R-charge assignment in such a way that all the fields have positive conformal weights. In the following we will set $R=\frac{1}{8}$. The product $ \prod_{\mathrm{\alpha \in all\,the\,roots}}$ runs over all the roots of $G_2$ and $\prod_{\mathrm{\rho_\Phi \in all\,the\,weights}}$ is including all the weights in a fundamental representation. 
The GNO charge $s$ \cite{Goddard:1976qe} is valued in a Cartan subalgebra  and quantized as
\begin{align}
s=s_1 H_1+ \sqrt{3} s_2 H_2,~~s_1 \ge 3s_2 , ~~ s_1,s_2 \in \mathbb{Z},
\end{align}
where we are restricting the summation of $(s_1,s_2)$ by using the Weyl reflections of $G_2$. $|\mathrm{Sym}|$ is a order of the Weyl group for the unbroken gauge group after the introduction of a GNO charge $(s_1,s_2)$.

\subsection*{$N_f=4$}

Since the superconformal indices contain negative powers of $x$ for $N_f \le 3$,
we start with the analysis of the superconformal indices from $N_f=4$. In order to have the positive R-charges for the chiral operators including the monopole operators, $R$ should be $0 < R < \frac{1}{4}$. Then the value $R=\frac{1}{8}$ is allowed. The theory with $N_f=4$ flavors is s-confining, so we have the dual description without gauge groups. We can compute the superconformal index on the electric theory and a magnetic one. We observe that these two indices exactly match. We first show the full conformal index for $N_f=4$.

\begin{align}
I (x,t)_{N_f=4}=1&+10 t^2 x^{1/4}+4 t^3 x^{3/8}+56 t^4 x^{1/2}+40 t^5 x^{5/8}+240 t^6 x^{3/4}+224 t^7 x^{7/8}\nonumber\\
&+\left(870 t^8+\frac{1}{t^8}\right) x+940 t^9 x^{9/8}+\left(2782 t^{10}+\frac{10}{t^6}\right) x^{5/4}+ \left(3280 t^{11}+4 t^{-5}  \right) x^{11/8} \nonumber\\
&+ \left(8055 t^{12}+\frac{55}{ t^4}\right) x^{3/2}+ \left(10008 t^{13}+\frac{36 }{t^{3}}  \right) x^{13/8}+  \left(21492 t^{14}+\frac{220}{ t^{2}} \right) x^{7/4}   \nonumber\\
& \quad + \left( 27536 t^{15}+\frac{180}{t}\right) x^{15/8} +\left(53495 t^{16}+\frac{1}{t^{16}}+698\right) x^2+ \cdots  
\end{align}
Next, we list the index for each GNO charge. This is obtained from the indices of the electric theory.
\begin{itemize}
\item GNO charge: (0,0)
\begin{align}
1&+10 t^2 x^{1/4}+4 t^3 x^{3/8}+56 t^4 x^{1/2}+40 t^5 x^{5/8}+240 t^6 x^{3/4}+224 t^7 x^{7/8}\nonumber\\
&+870 t^8 x+940 t^9 x^{9/8}+2782 t^{10} x^{5/4}+3280 t^{11} x^{11/8}+8055 t^{12} x^{3/2}\nonumber\\
&\quad +10008 t^{13} x^{13/8}+21492 t^{14} x^{7/4}+27536 t^{15} x^{15/8}+\left(53495 t^{16}-16\right) x^2+ \cdots 
\end{align}

\item GNO charge: (1,0)
\begin{align}
\frac{x}{t^8}&+\frac{10 x^{5/4}}{t^6}+\frac{4 x^{11/8}}{t^5}+\frac{55 x^{3/2}}{t^4}+\frac{36 x^{13/8}}{t^3} +\frac{220 x^{7/4}}{t^2}+\frac{180 x^{15/8}}{t}+714 x^2+\cdots
\end{align}

\item GNO charge: (2,0)
\begin{align}
\frac{x^2}{t^{16}}&+\frac{10 x^{9/4}}{t^{14}}+\frac{4 x^{19/8}}{t^{13}}+\frac{55 x^{5/2}}{t^{12}}+\frac{36 x^{21/8}}{t^{11}}+\frac{220 x^{11/4}}{t^{10}}+ \frac{180 x^{23/8}}{t^9} +\frac{714 x^3}{t^8} +\cdots 
\end{align}

\item GNO charge: (3,0)
\begin{align}
\frac{x^3}{t^{24}}+\frac{10 x^{13/4}}{t^{22}}+\frac{4
   x^{27/8}}{t^{21}}+\frac{55 x^{7/2}}{t^{20}}+\frac{36
   x^{29/8}}{t^{19}}+\frac{220 x^{15/4}}{t^{18}}+\frac{180
   x^{31/8}}{t^{17}}+\frac{714 x^4}{t^{16}}+\cdots 
\end{align}

\item GNO charge: (3,1)
\begin{align}
&\frac{x^8}{t^{32}}+\frac{4 x^{65/8}}{t^{31}}+\frac{10
   x^{33/4}}{t^{30}}+\frac{20 x^{67/8}}{t^{29}}+\frac{35
   x^{17/2}}{t^{28}}+\frac{56 x^{69/8}}{t^{27}}+\frac{84
   x^{35/4}}{t^{26}}+\frac{120 x^{71/8}}{t^{25}}+\frac{165
   x^9}{t^{24}}+\cdots 
\end{align}

\item GNO charge: (4,0)
\begin{align}
&\frac{x^4}{t^{32}}+\frac{10 x^{17/4}}{t^{30}}+\frac{4
   x^{35/8}}{t^{29}}+\frac{55 x^{9/2}}{t^{28}}+\frac{36
   x^{37/8}}{t^{27}}+\frac{220 x^{19/4}}{t^{26}}+\frac{180
   x^{39/8}}{t^{25}}+\frac{714 x^5}{t^{24}}+ \cdots
\end{align}

\item GNO charge: (4,1)
\begin{align}
\frac{x^9}{t^{40}}+\frac{4 x^{73/8}}{t^{39}}+\frac{10
   x^{37/4}}{t^{38}}+\frac{20 x^{75/8}}{t^{37}}+\frac{35
   x^{19/2}}{t^{36}}+\frac{56 x^{77/8}}{t^{35}}+\frac{84
   x^{39/4}}{t^{34}}+\frac{120 x^{79/8}}{t^{33}} +\cdots
\end{align}

\end{itemize}

We first explain low-lying operators in a sector with GNO charge $(0,0)$. Since the $G_2$ gauge group is unbroken in this sector, it is simple enough to understand the BPS operators.
The first contribution of unity is an identity operator with the GNO charges $(0,0)$. From the state-operator mapping it is denoted as $\ket{0,0}$. The second contribution $10 t^2 x^{1/4}$ is a meson $M_{ij}$ acting on $\ket{0,0}$. The third one $4 t^3 x^{3/8}$ is identified with $B^i \ket{0,0}$. The fourth term $56 t^4 x^{1/2}$ is from $F \ket{0,0}$ and $M_{ij} \otimes M_{kl} \ket{0,0}$, which are $\mathbf{1}+ \mathbf{20}+\mathbf{35}$ in an $SU(4)$ notation. The fifth term $40 t^5 x^{5/8}$ represents $M_{ij} \otimes B^k=\mathbf{4} +\mathbf{36}$. In this way we can find the chiral ring without monopole contributions.

Let us next consider a sector with the GNO charge $(1,0)$. In this case the gauge group is broken to $SU(2) \times U(1)$, so the chiral ring constructed on the state $ \ket{1,0}$ is modified from the previous case as in \cite{Bashkirov:2011vy, Kim:2013cma}.
From Table \ref{3dG2SQCD}, the monopole operator $Z$ which has a minimal magnetic charge appear as $t^{-8} x^1$ and this is consistent with the index above. The second and third contributions $\frac{10 x^{5/4}}{t^6}+\frac{4 x^{11/8}}{t^5}$ are identified with $M_{ij} \ket{1,0}$ and $B^i \ket{1,0}$ respectively. The fourth term $\frac{55 x^{3/2}}{t^4}$ only comes from $M_{ij} \otimes M_{kl} \ket{1,0}$ and the chiral ring does not have $F \ket{1,0}$. This is because we cannot construct the quartic baryons from the unbroken $SU(2)$ sector. The fifth term $\frac{36 x^{13/8}}{t^3}$ is also reduced because we cannot construct $\mathbf{4}$ which requires fourth order anti-symmetrization of the flavor indices and it is impossible. As the result, we only have a $\mathbf{36}$ representation.  
The sectors with GNO charges $(2,0)$, $(3,0)$ and $(4,0)$ are consistent with $(1,0)$ simply because the symmetry breaking pattern is the same.

For the sector with GNO charge $(3,1)$, we have to first notice that the gauge group is broken to $SU(2) \times U(1)$, where this $SU(2)$ is generated by the roots $\beta,-\beta$ and $\bld{\beta} \cdot \bld{H}$. Under this breaking, the fundamental representation is decomposed as
\begin{align}
\mathbf{7} \rightarrow \mathbf{2}_{1} +\mathbf{2}_{-1} +\mathbf{1}_2 + \mathbf{1}_{-2} +\mathbf{1}_0.
\end{align}
Therefore we can construct gauge invariant operators by acting the last component $\mathbf{1}_0$ on the monopole background with a GNO charge $(3,1)$. We do not have to combine two $Q$'s into $M$. 
The ground state $\ket{3,1}$ semi-classically corresponds to $Z^3 V_\alpha$ and the first excited state $\frac{4 x^{65/8}}{t^{31}}$ is $\mathbf{1}_0 \ket{3,1}$. The remaining parts are just given by symmetrizing $(\mathbf{1}_0)^{n}$ about the flavor indices.

In a sector with a GNO charge $(4,1)$, the gauge group is maximally broken to $U(1) \times U(1)$. In this broken phase the fundamental matters still supply the gauge singlet $\mathbf{1}_{(0,0)}$ so that we can construct the states
\begin{align}
\ket{4,1} & \Leftrightarrow \frac{x^9}{t^{40}},\\
\mathbf{1}_{(0,0)} \ket{4,1}  &  \Leftrightarrow  \frac{4 x^{73/8}}{t^{39}}, \\
\mathbf{1}_{(0,0)}\mathbf{1}_{(0,0)} \ket{4,1}  &  \Leftrightarrow  \frac{10x^{37/4}}{t^{38}},\\
&\vdots  \nonumber
\end{align}
where the flavor indices of $\mathbf{1}_{(0,0)}$ are symmetrized.

\vspace{0.6cm}
Finally we will list the superconformal indices for $N_f=5$ and $6$. The results are consistent with our finding that the Coulomb moduli space is labeled by $Z$.
\subsection*{$N_{f}=5$}

\begin{itemize}
\item GNO charge: (0,0)
\begin{align}
& 1+15 t^2 x^{1/4}+10 t^3 x^{3/8}+125 t^4 x^{1/2}+150 t^5
   x^{5/8}+805 t^6 x^{3/4}+1240 t^7 x^{7/8} 
   +8820 t^8 x \nonumber \\
  &  \quad +7570 t^9 
   x^{9/8} +21202 t^{10} x^{5/4}+37950 t^{11} x^{11/8}+91120
   t^{12} x^{3/2} 
   +164430 t^{13} x^{13/8}  \nonumber \\ 
  & \quad +355050 t^{14}x^{7/4} +634851 t^{15} x^{15/8}+\left(-25+1268710 t^{16}\right)
   x^2
   +\left(-50 t+2229135 t^{17}\right) x^{17/8} \nonumber \\
   &\quad +\left(-400
   t^2+4198290 t^{18}\right) x^{9/4}
   +\left(-950 t^3+7222165
   t^{19}\right) x^{19/8}  \nonumber\\
   &\quad +\left(-3825 t^4+12974178 t^{20}\right)
   x^{5/2} 
   +\left(-9225 t^5+21827235 t^{21}\right)
   x^{21/8}   \nonumber \\
   &\quad +\left(-27500 t^6+37715930 t^{22}\right)
   x^{11/4}
   +\left(-63350 t^7+62063820 t^{23}\right)
   x^{23/8}    \nonumber \\
   &\qquad +\left(-159750 t^8+103778515 t^{24}\right)
   x^3
   +\left(-347425 t^9+167175552 t^{25}\right) x^{25/8}+\cdots
\end{align}
\item GNO charge: (1,0)
\begin{align}
\frac{x^{11/4}}{t^{10}}&+\frac{15 x^3}{t^8}+\frac{10
   x^{25/8}}{t^7}+\frac{120 x^{13/4}}{t^6}+\frac{126
   x^{27/8}}{t^5}+\frac{680 x^{7/2}}{t^4}\nonumber\\
   &\qquad \qquad+\frac{855
   x^{29/8}}{t^3}+\frac{3045 x^{15/4}}{t^2}+\frac{4145
   x^{31/8}}{t}+11427 x^4 +\cdots
\end{align}

\item GNO charge: (2,0)  
\begin{align}
\frac{x^{11/2}}{t^{20}}&+\frac{15 x^{23/4}}{t^{18}}+\frac{10
   x^{47/8}}{t^{17}}+\frac{120 x^6}{t^{16}}+\frac{126
   x^{49/8}}{t^{15}}+\frac{680 x^{25/4}}{t^{14}}+\frac{855
   x^{51/8}}{t^{13}}\nonumber\\
   &\qquad +\frac{3045 x^{13/2}}{t^{12}}+\frac{4145
   x^{53/8}}{t^{11}}+\frac{11427 x^{27/4}}{t^{10}}+\frac{16080
   x^{55/8}}{t^9}+\frac{37310 x^7}{t^8} +\cdots
\end{align}

\item GNO charge: (3,0)
\begin{align}
\frac{x^{33/4}}{t^{30}}+\frac{15 x^{17/2}}{t^{28}}+\frac{10
   x^{69/8}}{t^{27}}+\frac{120 x^{35/4}}{t^{26}}+\frac{126
   x^{71/8}}{t^{25}}+\frac{680 x^9}{t^{24}}+\frac{855
   x^{73/8}}{t^{23}}+\frac{3045 x^{37/4}}{t^{22}} +\cdots
\end{align}

\item GNO charge: (3,1)
\begin{align}
&\frac{x^{15}}{t^{40}}
+\frac{5 x^{121/8}}{t^{39}}+\frac{15
   x^{61/4}}{t^{38}}+\frac{35 x^{123/8}}{t^{37}}+\frac{70
   x^{31/2}}{t^{36}}  \nonumber \\
  & \qquad \qquad \qquad \quad+\frac{126 x^{125/8}}{t^{35}}
   +\frac{210
   x^{63/4}}{t^{34}}+\frac{330 x^{127/8}}{t^{33}}+\frac{495
   x^{16}}{t^{32}}+\cdots 
\end{align}

\item GNO charge: (4,0)
\begin{align}
\frac{x^{11}}{t^{40}}&+\frac{15 x^{45/4}}{t^{38}}+\frac{10
   x^{91/8}}{t^{37}}+\frac{120 x^{23/2}}{t^{36}}+\frac{126
   x^{93/8}}{t^{35}}+\frac{680 x^{47/4}}{t^{34}}\nonumber\\
   &\qquad  \qquad \quad +\frac{855x^{95/8}}{t^{33}}+\frac{3045 x^{12}}{t^{32}}+\frac{4145
   x^{97/8}}{t^{31}}+\frac{11427 x^{49/4}}{t^{30}}+\cdots 
\end{align}
\end{itemize}

\subsection*{$N_f=6$}
\begin{itemize}
\item GNO charge: (0,0)
\begin{align}
\begin{autobreak}
1
+21 t^2 x^{1/4}
+20 t^3 x^{3/8}
+246 t^4 x^{1/2}
+420 t^5x^{5/8}
+2261 t^6 x^{3/4} 
+4830 t^7 x^{7/8}
+17766 t^8 x
+40740 t^9  x^{9/8}
+121569 t^{10} x^{5/4}  
+280140 t^{11} x^{11/8}
+733194 t^{12} x^{3/2}
+1651440 t^{13} x^{13/8}
+3946974 t^{14} x^{7/4} 
 +8597092 t^{15} x^{15/8}
 +\left(-36+19195449 t^{16}\right) x^2 
 +\left(-90t+40315392 t^{17}\right) x^{17/8}
   +  \left(-876t^2+85267989 t^{18}\right) x^{9/4}
   + \left(-2610 t^3+172772712 t^{19}\right) x^{19/8}
   +  \left(-12636t^4+349323471 t^{20}\right) x^{5/2}
   +    \left(-38100 t^5+684175032 t^{21}\right) x^{21/8}
   +\left(-135576t^6+ 1330939701 t^{22}\right) x^{11/4}
  \quad + \left(-386820t^7+2525733672 t^{23}\right) x^{23/8}
  +\left(-1160376t^8+4750153876 t^{24}\right) x^3+\cdots
   \end{autobreak}
\end{align}

\item GNO charge: (1,0)
\begin{align}
&\frac{x^{9/2}}{t^{12}}+\frac{21 x^{19/4}}{t^{10}}+\frac{20
   x^{39/8}}{t^9}+\frac{231 x^5}{t^8}+\frac{336
   x^{41/8}}{t^7}+\frac{1771 x^{21/4}}{t^6} +\frac{2976
   x^{43/8}}{t^5}  \nonumber  \\
 &\qquad  +\frac{10521 x^{11/2}}{t^4} 
   +\frac{18480
   x^{45/8}}{t^3}+\frac{51309 x^{23/4}}{t^2} +\frac{90300
   x^{47/8}}{t}+213479 x^6+\cdots  
\end{align}
\item GNO charge: (2,0)
\begin{align}
\frac{x^9}{t^{24}}&+\frac{21 x^{37/4}}{t^{22}}+\frac{20
  x^{75/8}}{t^{21}}+\frac{231 x^{19/2}}{t^{20}}+\frac{336
   x^{77/8}}{t^{19}}+\frac{1771 x^{39/4}}{t^{18}} +\frac{2976
   x^{79/8}}{t^{17}}+\frac{10521 x^{10}}{t^{16}} +\cdots
\end{align}
\item GNO charge: (3,0)
\begin{multline}
\frac{x^{27/2}}{t^{36}}+\frac{21 x^{55/4}}{t^{34}}+\frac{20
   x^{111/8}}{t^{33}}+\frac{231 x^{14}}{t^{32}}+\frac{336
   x^{113/8}}{t^{31}}+\frac{1771 x^{57/4}}{t^{30}}+\frac{2976
   x^{115/8}}{t^{29}}+\frac{10521 x^{29/2}}{t^{28}} +\cdots
\end{multline}

\item GNO charge: (3,1)
\begin{multline}
\frac{x^{22}}{t^{48}}+\frac{6 x^{177/8}}{t^{47}}+\frac{21
   x^{89/4}}{t^{46}}+\frac{56 x^{179/8}}{t^{45}}+\frac{126
   x^{45/2}}{t^{44}}  \\ +\frac{252 x^{181/8}}{t^{43}}+\frac{462
   x^{91/4}}{t^{42}}+\frac{792 x^{183/8}}{t^{41}}+\frac{1287
   x^{23}}{t^{40}} +\cdots
\end{multline}

\end{itemize}

\section{Summary and Discussion}
We investigated the 3d $\mathcal{N}=2$ supersymmetric $G_2$ gauge theory with (and without) fundamental matters. We found that the Coulomb branch of the moduli space of vacua is classically two-dimensional but the monopole-instantons generate the runaway-type superpotential and make the one-dimensional subspace massive. As the result, the quantum Coulomb moduli space is one-complex dimensional and we provided the proper monopole operator describing it. We introduced the chiral superfields in a fundamental representation and discussed that this one-dimensional direction remains after including the matters and their zero-modes.
We also found that there are various phases depending on the number of the fundamentals. For $N_f \le 2$, we have no supersymmetric vacuum. For $N_f=3$, the Coulomb and Higgs moduli are quantum-mechanically merged and relating the weak- and strong-coupling regions. For $N_f=4$, we found the s-confinement phase where the dual description is given by only gauge-singlet chiral superfields. As an independent check of our analysis, we calculated the superconformal indices and confirmed that the Coulomb branch is indeed parametrized by a $Z$ field and observed the correct low-lying BPS operators.

In  this paper we have shown the existence of the one-dimensional Coulomb branch, so it is possible to calculate a Hilbert series a la \cite{Cremonesi:2015dja, Hanany:2015via, Benvenuti:2006qr, Feng:2007ur, Pouliot:1998yv}. Hilbert series basically counts the holomorphic (gauge invariant) operators in a theory. Then we can study another aspect of the $G_2$ gauge theory and check the validity of our analysis. It is also interesting to consider the $G_2$ Chern-Simons theory.

A simple generalization of this work would be to study other exceptional groups in a framework of a 3d $\mathcal{N}=2$ supersymmetry. In 4d, such theories do not have any s-confinement phases but have some quantum-deformed moduli spaces (see for instance \cite{Ramond:1996ku, Distler:1996ub, Karch:1997jp, Cho:1997am, Pouliot:2001iw, Csaki:1997aw}). Naively we expect that this is also the case in 3d. However, when connecting the physics between 3d and 4d, it is often the case where the s-confinement phase in 3d is de-compactified to the quantum deformed moduli space in 4d via the KK-monopole superpotential. So we can expect that some s-confinement phases might emerge in 3d for $F_4,E_6,E_7$ and $E_8$ being different from the 4d cases. It is also interesting to study the 3d Seiberg duality for those exceptional groups.

In this paper we only included the fundamental matters. So it is interesting to add some matter chiral superfields in various representations. In 4d if we include many matter fields, the theory is no longer asymptotically free. But in 3d the gauge coupling is a relevant interaction. Then it is interesting to study those cases. The possible matters would be adjoint. When studying those theories, the Coulomb branch becomes complicated to study because the Coulomb branch is no longer one-dimensional. Therefore it is a first attempt to consider the (adjoint) matter with some superpotential. The presence of the superpotential caves the chiral ring and would simplify its analysis.

We could not find any Seiberg dual description for $N_f \ge 5$ where the ``effective'' superpotential had the singularities at the origin of the moduli space. This is implicitly telling us the presence of a magnetic gauge group and dual quarks. In 4d, the $G_2$ Seiberg dual is known in \cite{Pouliot:1995zc} and we reviewed it in Section 2. Naively speaking we can derive the corresponding 3d Seiberg dual by dimensionally reducing the 4d electric and magnetic theories respectively. This method was studied in \cite{Aharony:2013dha, Aharony:2013kma}, where those authors claimed that in reducing the 4d Seiberg dual pairs to 3d, it is important to take into account the non-perturbative effects from the twisted instantons (KK-monopoles) and carefully to take some low-energy limit on both sides. In our case of $G_2$, we can easily find the electric theory on a circle. This is just including the superpotential $W= \eta Z$. On the magnetic side, however, it is difficult to study the full Coulomb branch structure and also difficult to derive the KK-monopole generated potential. More concretely we are not understanding dimensions of the Coulomb (quantum) moduli. For example, the Coulomb branch operator corresponding to 
\begin{align}
Y_{SU(N_f-3)} \leftrightarrow \begin{pmatrix}
 \phi &  &  &&\\
 &0&& &\\
  & &\ddots && \\
  &&&0& \\
&  &  &&  -\phi
\end{pmatrix}
\end{align}
is not gauge invariant because the magnetic theory is ``chiral'' in a four-dimensional sense, which includes only anti-fundamentals and not fundamentals. We can construct a dressed monopole operator by multiplying the chiral superfields a la \cite{Csaki:2014cwa, Amariti:2015kha}. In the present case, we find that the following dressed monopole can be defined.
\begin{align}
Y_{dressed} := Y_{SU(N_f-3)} s^{N_f-5} 
\end{align}
This is quite plausible because the superpotential on the magnetic theory contains $W \ni \mathrm{det} \, s= s^{N_f-3}$ and such a dressed monopole would be generated by absorbing the fermion zero-modes from the symmetric matter. We also found that the non-perturbative superpotential
\begin{align}
W= \tilde{\eta} Y_{dressed}
\end{align}
would be generated via the KK-monopole and dressing effects. This superpotential is consistent with all the (spurious) symmetries. But we do not understand whether any other Coulomb branch directions quantum mechanically remain massless and whether other types of KK-monopole superpotential might be generated or not. We have to also take a 3d limit in order to turn off the electric superpotential $W= \eta Z$. This can be achieved by introducing real masses by background gauging the flavor symmetry $SU(N_f)$. On the magnetic side, this deformation would lead to the higgsing of the dual gauge group. Under this higgsing the  Coulomb branch operator $Y_{dressed}$ is non-trivially transformed and additional Coulomb branch operators would also emerge. We have to rewrite the magnetic superpotential in this new set of monopole operators. This is highly non-trivial and we could not find any natural dual description.

It remains important to study a 3d $\mathcal{N}=2$ $Spin(7)$ theory with spinorial representations $\mathbf{8}$ since the $G_2$ gauge theory comes from this by higgsing the $Spin(7)$ gauge group via a vev of the spinorial scalar field. Although we could not find a 3d $G_2$ dual description from the 4d $G_2$ Seiberg duality, it is quite plausible that we can find the $G_2$ duals after constructing the 3d Seiberg duality for $Spin(7)$. We will come back to this problem and near future we would like to address some progresses on this direction.


\section*{Acknowledgments}

We are grateful to Antonio Amariti for valuable discussions. K. N. would also like to thank Uwe-Jens Wiese and Susanne Reffert for helpful discussions and comments.  The work of K.N. and Y.S. is supported by the Swiss National Science Foundation (SNF) under grant number PP00P2\_157571/1.

\appendix

\section{$G_2$ notations}
Here we summarize notations for the $G_2$ group and its Lie algebra, which we have used in this paper. 
For the details of the $G_2$ algebra and its representations, for example, see \cite{Behrends:1962zz, Slansky:1981yr}.

\subsection{Group invariants}
The $G_2$ is a smallest exceptional Lie group with a trivial center. It has rank 2 and dimension 14. 
$G_2$ has maximal subgroups $SU(3)$ and $SU(2) \times SU(2)$.

The Dynkin index $T_{\mathbf{r}}$ is defined as a constant appearing in $\mathrm{Tr} \, ( t^a t^b )= T_{\mathbf{r}} \, \delta^{ab}$. For fundamental and adjoint representations it is given by
\begin{align}
T_\mathbf{7} =1, ~~~T_{\mathbf{Adj.}} =4.
\end{align}
A one-loop beta function in a 4d $\mathcal{N}=1$ SQCD is given by
\begin{align}
\beta(g) =-\frac{g^3}{16 \pi^2}b,~~~b=  3T_{\mathbf{Adj.}} -\sum_i T_{\mathbf{r_i}} ,
\end{align}
where $3T_{\mathbf{Adj.}}$ is a contribution from the vector superfield and the other is from chiral superfields with representations $\mathbf{r}_i$.

It is useful to enumerate the group invariant tensors for the group $G_2$. We have two invariant tensors. The first one is a Kronecker delta symbol $\delta_{ab}$ where $a,b=1,\cdots,7$. The second one is a totally anti-symmetric tensor $f_{abc}$. In addition to these tensors we can construct the fourth order totally antisymmetric tensor $\tilde{f}_{abcd} := f_{e[ab} f_{cd]e}$, which is also expressed by the dual of $f_{abc}$.

The $G_{2}$ gauge invariants can be constructed by contracting these invariant tensors
with the fundamental quarks. The composite fields are thus given by
\begin{align}
M_{ij} &=\delta_{ab}\,Q^{a}{}_{i}Q^{b}{}_{j}\,,\nonumber\\
B_{ijk} &=\frac{1}{3!}f_{abc} \,Q^{a}{}_{i}Q^{b}{}_{j}Q^{c}{}_k\,\,,\nonumber\\
F_{ijkl} &= \frac{1}{4!} \tilde{f}_{abcd} \,Q^{a}{}_i Q^{b}{}_jQ^{c}{}_kQ^{d}{}_l
\end{align}
%

\subsection{Representations of $G_{2}$}
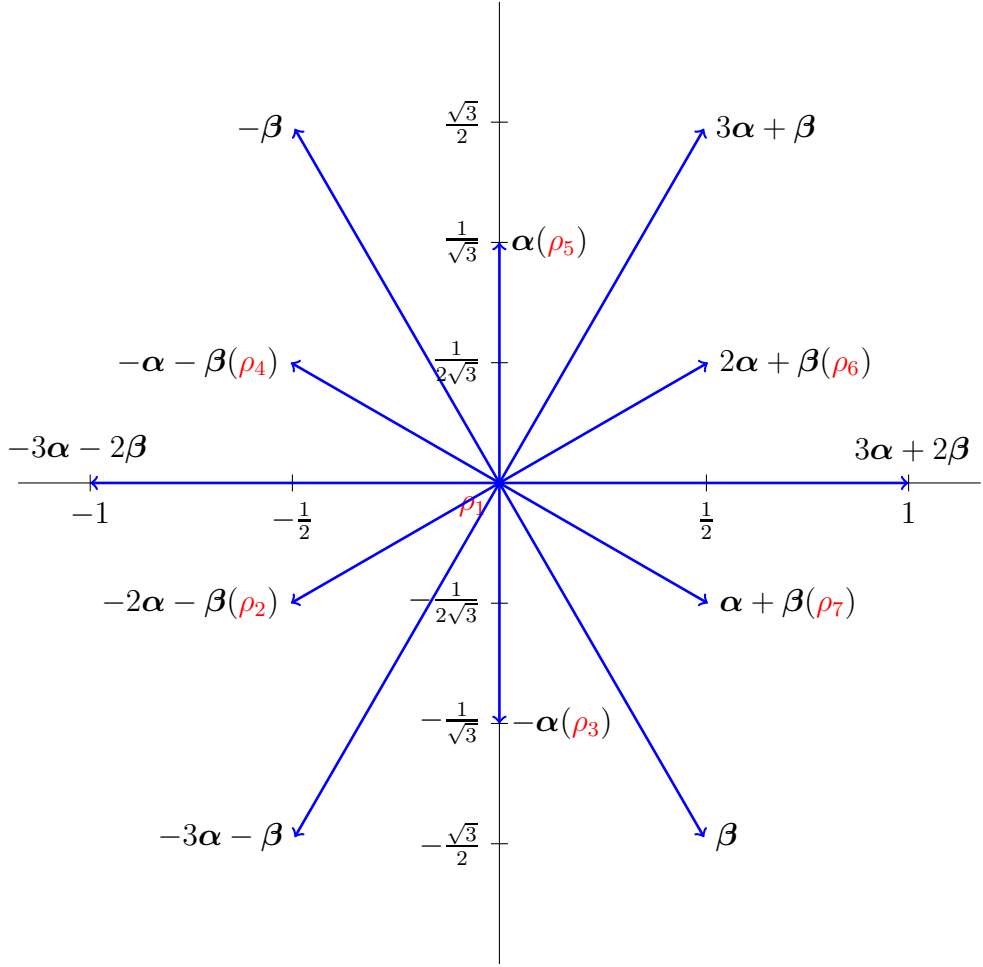
\begin{figure}[H]
\begin{tikzpicture}
         [scale=3.2,
     kutje/.style={circle,draw=black,fill=white, thick, inner sep=0pt, minimum size=7mm}
   ]
    \coordinate (a) at (0,0);
    \coordinate (b0) at (0,1);
    \coordinate (b1) at ($ (a)! 1! 60:(b0) $);
    \coordinate (b2) at ($ (a)! 1! 120:(b0) $);
    \coordinate (b3) at ($ (a)! 1! 180:(b0) $);
    \coordinate (b4) at ($ (a)! 1! 240:(b0) $);
    \coordinate (b5) at ($ (a)! 1! -60:(b0) $);
    \coordinate (c0) at ($ (a)! 1.7! 150:(b0) $);
    \coordinate (c1) at ($ (a)! 1.7! 30:(b0) $);
    \coordinate (c2) at ($ (a)! 1.7! 90:(b0) $);
    \coordinate (c3) at ($ (a)! 1.7! -150:(b0) $);
    \coordinate (c4) at ($ (a)! 1.7! -30:(b0) $);
    \coordinate (c5) at ($ (a)! 1.7! -90:(b0) $);
    \draw (0,-2) -- (0,2);
    \draw (-2,0) -- (2,0);
    \draw[color=black] (a) node[red=black, left=10pt, below=1pt]{$\rho_{1}$};
    \draw[color=blue, line width=1pt,->] (a) -- (b0) node[color=black, right] {$\bld{\alpha}(\color{red}\rho_{5}\color{black})$};
    \draw[color=blue, line width=1pt,->] (a) -- (b1) node[color=black, left]{$-\bld{\alpha}-\bld{\beta}(\color{red}\rho_{4}\color{black})$};
    \draw[color=blue, line width=1pt,->] (a) -- (b2) node[color=black, left]{$-2\bld{\alpha}-\bld{\beta}(\color{red}\rho_{2}\color{black})$};
    \draw[color=blue, line width=1pt,->] (a) -- (b3) node[color=black, right]{$-\bld{\alpha}(\color{red}\rho_{3}\color{black})$};
    \draw[color=blue, line width=1pt,->] (a) -- (b4) node[color=black, right]{$\bld{\alpha}+\bld{\beta}(\color{red}\rho_{7}\color{black})$};
    \draw[color=blue, line width=1pt,->] (a) -- (b5) node[color=black, right]{$2\bld{\alpha}+\bld{\beta}(\color{red}\rho_{6}\color{black})$};
    \draw[color=blue, line width=1pt,->] (a) -- (c0) node[color=black, left] {$-3\bld{\alpha}-\bld{\beta}$};
    \draw[color=blue, line width=1pt,->] (a) -- (c1) node[color=black, left] {$-\bld{\beta}$};
    \draw[color=blue, line width=1pt,->] (a) -- (c2) node[color=black, left=5pt, above=3pt]{$-3\bld{\alpha}-2\bld{\beta}$};
    \draw[color=blue, line width=1pt,->] (a) -- (c3) node[color=black, right]{$\bld{\beta}$};
    \draw[color=blue, line width=1pt,->] (a) -- (c4) node[color=black, right]{$3\bld{\alpha}+\bld{\beta}$};
    \draw[color=blue, line width=1pt,->] (a) -- (c5) node[color=black, above=12pt, right=-25pt]{$3\bld{\alpha}+2\bld{\beta}$};
    \foreach \x/\xtext in {-1.7/-1, -0.86/-\frac{1}{2}, 0.86/\frac{1}{2},1.7/1}
    \draw (\x cm,1pt) -- (\x cm,-1pt) node[anchor=north] {$\xtext$};
\foreach \y/\ytext in {-1.5/-\frac{\sqrt{3}}{2}, -1/-\frac{1}{\sqrt{3}}, -0.5/-\frac{1}{2\sqrt{3}} , 0.5/ \frac{1}{2\sqrt{3}}, 1/\frac{1}{\sqrt{3}}, 1.5/\frac{\sqrt{3}}{2}}
    \draw (1pt,\y cm) -- (-1pt,\y cm) node[anchor=east] {$\ytext$};
\end{tikzpicture}
\centering
\caption{The $G_{2}$ roots and weights of fundamental representations}
  \label{fig:G2Roots}
\end{figure}

We follow the notation of the Lie algebra for $G_2$ used in \cite{Ruben} although we are relabeling the names. The adjoint representation is represented by $7 \times 7$ matrices with $14$ generators, which are decomposed into two Cartan matrices and $12$ raising and lowering operators.
In this representation, the fundamental representation with $7$ dimensions are taking a 7 dimensional column vector and the matrices below naturally act on the column vector.  
The explicit parametrization for the Cartan subalgebra and 12 roots are as follows. 
\begin{align}
X_{\alpha} &=
\begin{pmatrix}
 0 & 0 & 2 & 0 & 0 & 0 & 0\\
 0 & 0 & 0 & 0 & 0 & 0 & 0\\
 0 & 0 & 0 & 0 & 0 & 0 & 0\\
 0 & -1 & 0 & 0 & 0 & 0 & 0\\
 1 & 0 & 0 & 0 & 0 & 0 & 0\\
 0 & 0 & 0 & 0 & 0 & 0 & -1\\
 0 & 0 & 0 & 0 & 0 & 0 & 0
\end{pmatrix}\,,\quad
&X_{-\alpha} = 
\begin{pmatrix}
 0 & 0 & 0 & 0 & 2 & 0 & 0\\
 0 & 0 & 0 & -1 & 0 & 0 & 0\\
 1 & 0 & 0 & 0 & 0 & 0 & 0\\
 0 & 0 & 0 & 0 & 0 & 0 & 0\\
 0 & 0 & 0 & 0 & 0 & 0 & 0\\
 0 & 0 & 0 & 0 & 0 & 0 & 0\\
 0 & 0 & 0 & 0 & 0 & -1 & 0
\end{pmatrix}\,,
\end{align}
\begin{align}
X_{\beta} &=
\begin{pmatrix}
 0 & 0 & 0 & 0 & 0 & 0 & 0\\
 0 & 0 & 0 & 0 & 0 & 0 & 0\\
 0 & 0 & 0 & -1 & 0 & 0 & 0\\
 0 & 0 & 0 & 0 & 0 & 0 & 0\\
 0 & 0 & 0 & 0 & 0 & 0 & 0\\
 0 & 0 & 0 & 0 & 0 & 0 & 0\\
 0 & 0 & 0 & 0 & 1 & 0 & 0
\end{pmatrix}\,,\quad
& X_{-\beta} =
\begin{pmatrix}
 0 & 0 & 0 & 0 & 0 & 0 & 0\\
 0 & 0 & 0 & 0 & 0 & 0 & 0\\
 0 & 0 & 0 & 0 & 0 & 0 & 0\\
 0 & 0 & 1 & 0 & 0 & 0 & 0\\
 0 & 0 & 0 & 0 & 0 & 0 & -1\\
 0 & 0 & 0 & 0 & 0 & 0 & 0\\
 0 & 0 & 0 & 0 & 0 & 0 & 0
\end{pmatrix}\,,
\end{align}
\begin{align}
X_{\alpha+\beta} &= 
\begin{pmatrix}
 0 & 0 & 0 & 2 & 0 & 0 & 0\\
 0 & 0 & 0 & 0 & 0 & 0 & 0\\
 0 & 1 & 0 & 0 & 0 & 0 & 0\\
 0 & 0 & 0 & 0 & 0 & 0 & 0\\
 0 & 0 & 0 & 0 & 0 & 0 & 0\\
 0 & 0 & 0 & 0 & 1 & 0 & 0\\
 1 & 0 & 0 & 0 & 0 & 0 & 0
\end{pmatrix}\,,\quad
& X_{-\alpha-\beta} = 
\begin{pmatrix}
  0 & 0 & 0 & 0 & 0 & 0 & 2\\
 0 & 0 & 1 & 0 & 0 & 0 & 0\\
 0 & 0 & 0 & 0 & 0 & 0 & 0\\
 1 & 0 & 0 & 0 & 0 & 0 & 0\\
 0 & 0 & 0 & 0 & 0 & 1 & 0\\
 0 & 0 & 0 & 0 & 0 & 0 & 0\\
 0 & 0 & 0 & 0 & 0 & 0 & 0
\end{pmatrix}\,,
\end{align}
\begin{align}
X_{2\alpha+\beta}&=
\begin{pmatrix}
 0 & -2 & 0 & 0 & 0 & 0 & 0\\
 0 & 0 & 0 & 0 & 0 & 0 & 0\\
 0 & 0 & 0 & 0 & 0 & 0 & 0\\
 0 & 0 & 0 & 0 & 0 & 0 & 0\\
 0 & 0 & 0 & -1 & 0 & 0 & 0\\
 1 & 0 & 0 & 0 & 0 & 0 & 0\\
 0 & 0 & 1 & 0 & 0 & 0 & 0
\end{pmatrix}\,,\quad
& X_{-2\alpha-\beta} =
\begin{pmatrix}
 0 & 0 & 0 & 0 & 0 & -2 & 0\\
 1 & 0 & 0 & 0 & 0 & 0 & 0\\
 0 & 0 & 0 & 0 & 0 & 0 & -1\\
 0 & 0 & 0 & 0 & 1 & 0 & 0\\
 0 & 0 & 0 & 0 & 0 & 0 & 0\\
 0 & 0 & 0 & 0 & 0 & 0 & 0\\
 0 & 0 & 0 & 0 & 0 & 0 & 0
\end{pmatrix}\,,
\end{align}
\begin{align}
X_{3\alpha+\beta}&=
\begin{pmatrix}
 0 & 0 & 0 & 0 & 0 & 0 & 0\\
 0 & 0 & 0 & 0 & 0 & 0 & 0\\
 0 & 0 & 0 & 0 & 0 & 0 & 0\\
 0 & 0 & 0 & 0 & 0 & 0 & 0\\
 0 & 1 & 0 & 0 & 0 & 0 & 0\\
 0 & 0 & 1 & 0 & 0 & 0 & 0\\
 0 & 0 & 0 & 0 & 0 & 0 & 0
\end{pmatrix}\,,\quad
& X_{-3\alpha-\beta} =
\begin{pmatrix}
 0 & 0 & 0 & 0 & 0 & 0 & 0\\
 0 & 0 & 0 & 0 & 0 & -1 & 0\\
 0 & 0 & 0 & 0 & 0 & 0 & 1\\
 0 & 0 & 0 & 0 & 0 & 0 & 0\\
 0 & 0 & 0 & 0 & 0 & 0 & 0\\
 0 & 1 & 0 & 0 & 0 & 0 & 0\\
 0 & 0 & -1 & 0 & 0 & 0 & 0
\end{pmatrix}\,,
\end{align}
\begin{align}
X_{3\alpha+2\beta} &=  
\begin{pmatrix}
 0 & 0 & 0 & 0 & 0 & 0 & 0\\
 0 & 0 & -1 & 0 & 0 & 0 & 0\\
 0 & 1 & 0 & 0 & 0 & 0 & 0\\
 0 & 0 & 0 & 0 & -1 & 0 & 0\\
 0 & 0 & 0 & 1 & 0 & 0 & 0\\
 0 & 0 & 0 & 1 & 0 & 0 & 0\\
 0 & 1 & 0 & 0 & 0 & 0 & 0
\end{pmatrix}\,,\quad
& X_{-3\alpha-2\beta} =  
\begin{pmatrix}
 0 & 0 & -1 & 0 & 0 & 0 & 0\\
 0 & 0 & 0 & 0 & 0 & 0 & 0\\
 1 & 0 & 0 & 0 & 0 & 0 & 0\\
 0 & 0 & 0 & 0 & 0 & -1 & 0\\
 0 & 0 & 0 & 0 & 0 & 0 & 0\\
 0 & 0 & 0 & 1 & 0 & 0 & 0\\
 0 & 0 & 0 & 0 & 0 & 0 & 0
\end{pmatrix}\,,
\end{align}
\begin{align}
H_{1} &=\frac{1}{2}
\begin{pmatrix}
 0 & 0 & 0 & 0 & 0 & 0 & 0\\
 0 & -1 & 0 & 0 & 0 & 0 & 0\\
 0 & 0 & 0 & 0 & 0 & 0 & 0\\
 0 & 0 & 0 & -1 & 0 & 0 & 0\\
 0 & 0 & 0 & 0 & 0 & 0 & 0\\
 0 & 0 & 0 & 0 & 0 & 1 & 0\\
 0 & 0 & 0 & 0 & 0 & 0 & 1
\end{pmatrix}\,,\quad 
& H_{2} = \frac{1}{2\sqrt{3}}
\begin{pmatrix}
 0 & 0 & 0 & 0 & 0 & 0 & 0\\
 0 & -1 & 0 & 0 & 0 & 0 & 0\\
 0 & 0 & -2 & 0 & 0 & 0 & 0\\
 0 & 0 & 0 & 1 & 0 & 0 & 0\\
 0 & 0 & 0 & 0 & 2 & 0 & 0\\
 0 & 0 & 0 & 0 & 0 & 1 & 0\\
 0 & 0 & 0 & 0 & 0 & 0 & -1
\end{pmatrix}\,.
\end{align}
%
Note that those generators in Cartan subalgebra are normalized such that

\begin{equation}
\mathrm{Tr} \, H_{1}H_{1} = \mathrm{Tr} \,H_{2}H_{2} = 1\,,\quad \mathrm{Tr} \,H_{1}H_{2} = 0\,.
\end{equation}
The two simple roots are expressed in a $(H_1,H_2)-$plane as
\begin{align}
\bld{\alpha}(H) = \left(0\,, \frac{1}{\sqrt{3}}\right)\,,\quad  \bld{\beta}(H) = \left(\frac{1}{2}\,,-\frac{\sqrt{3}}{2}\right)
\end{align}
and the other positive roots are $\bld{\alpha}+\bld{\beta},~2\bld{\alpha}+\bld{\beta},~3\bld{\alpha}+\bld{\beta}$ and $3\bld{\alpha}+2\bld{\beta}$.
For a fundamental representation, we chose a following set of weights:
\begin{align}
\rho_{i} \cong {}^{t}(0,\dots,\stackrel{i\mathchar`-\mathrm{th}}{1},\dots,0)\,,\quad i=1,\dots,7\,,
\end{align}
which can be parametrized on the $(H_{1},H_{2})$-plane as
\begin{align}
{\bf \rho}_{i}(H) = \biggl( (H_{1})_{ii}, (H_{2})_{ii} \biggr)\,,\quad i:\ \mathrm{not\ summed}
\end{align}
The $G_{2}$ root system and a weight diagram of a fundamental representation are depicted in Figure \ref{fig:G2Roots}.

\bibliographystyle{ieeetr}
\bibliography{3dG2}

\begin{thebibliography}{10}

\bibitem{Seiberg:1994bz}
N.~Seiberg, ``{Exact results on the space of vacua of four-dimensional SUSY
  gauge theories},'' {\em Phys. Rev.}, vol.~D49, pp.~6857--6863, 1994.

\bibitem{Seiberg:1994pq}
N.~Seiberg, ``{Electric - magnetic duality in supersymmetric nonAbelian gauge
  theories},'' {\em Nucl. Phys.}, vol.~B435, pp.~129--146, 1995.

\bibitem{Seiberg:1997vw}
N.~Seiberg, ``{The Power of duality: Exact results in 4-D SUSY field theory},''
  {\em Int. J. Mod. Phys.}, vol.~A16, pp.~4365--4376, 2001.
\newblock [Prog. Theor. Phys. Suppl.123,337(1996)].

\bibitem{Holland:2002vk}
K.~Holland, P.~Minkowski, M.~Pepe, and U.~J. Wiese, ``{Confinement without a
  center: The Exceptional group G(2)},'' {\em Nucl. Phys. Proc. Suppl.},
  vol.~119, pp.~652--654, 2003.
\newblock [,652(2002)].

\bibitem{Holland:2003jy}
K.~Holland, P.~Minkowski, M.~Pepe, and U.~J. Wiese, ``{Exceptional confinement
  in G(2) gauge theory},'' {\em Nucl. Phys.}, vol.~B668, pp.~207--236, 2003.

\bibitem{Pepe:2006er}
M.~Pepe and U.~J. Wiese, ``{Exceptional Deconfinement in G(2) Gauge Theory},''
  {\em Nucl. Phys.}, vol.~B768, pp.~21--37, 2007.

\bibitem{Cossu:2007dk}
G.~Cossu, M.~D'Elia, A.~Di~Giacomo, B.~Lucini, and C.~Pica, ``{G(2) gauge
  theory at finite temperature},'' {\em JHEP}, vol.~10, p.~100, 2007.

\bibitem{Wellegehausen:2010ai}
B.~H. Wellegehausen, A.~Wipf, and C.~Wozar, ``{Casimir Scaling and String
  Breaking in G(2) Gluodynamics},'' {\em Phys. Rev.}, vol.~D83, p.~016001,
  2011.

\bibitem{Bruno:2014rxa}
M.~Bruno, M.~Caselle, M.~Panero, and R.~Pellegrini, ``{Exceptional
  thermodynamics: the equation of state of G$_{2}$ gauge theory},'' {\em JHEP},
  vol.~03, p.~057, 2015.

\bibitem{Poppitz:2012nz}
E.~Poppitz, T.~Schafer, and M.~Unsal, ``{Universal mechanism of
  (semi-classical) deconfinement and theta-dependence for all simple groups},''
  {\em JHEP}, vol.~03, p.~087, 2013.

\bibitem{Alishahiha:1995wm}
M.~Alishahiha, F.~Ardalan, and F.~Mansouri, ``{The Moduli space of the
  supersymmetric G(2) Yang-Mills theory},'' {\em Phys. Lett.}, vol.~B381,
  pp.~446--450, 1996.

\bibitem{Landsteiner:1996ut}
K.~Landsteiner, J.~M. Pierre, and S.~B. Giddings, ``{On the moduli space of N=2
  supersymmetric G(2) gauge theory},'' {\em Phys. Rev.}, vol.~D55,
  pp.~2367--2372, 1997.

\bibitem{Pesando:1995bq}
I.~Pesando, ``{Exact results for the supersymmetric G(2) gauge theories},''
  {\em Mod. Phys. Lett.}, vol.~A10, pp.~1871--1886, 1995.

\bibitem{Giddings:1995ns}
S.~B. Giddings and J.~M. Pierre, ``{Some exact results in supersymmetric
  theories based on exceptional groups},'' {\em Phys. Rev.}, vol.~D52,
  pp.~6065--6073, 1995.

\bibitem{Smilga:1998em}
A.~V. Smilga, ``{6+1 vacua in supersymmetric QCD with G(2) gauge group},'' {\em
  Phys. Rev.}, vol.~D58, p.~105014, 1998.

\bibitem{Davies:2000nw}
N.~M. Davies, T.~J. Hollowood, and V.~V. Khoze, ``{Monopoles, affine algebras
  and the gluino condensate},'' {\em J. Math. Phys.}, vol.~44, pp.~3640--3656,
  2003.

\bibitem{Alishahiha:2003hj}
M.~Alishahiha, J.~de~Boer, A.~E. Mosaffa, and J.~Wijnhout, ``{N=1 G(2) SYM
  theory and compactification to three-dimensions},'' {\em JHEP}, vol.~09,
  p.~066, 2003.

\bibitem{Saito:2007ah}
O.~Saito, ``{The Glueball superpotential for G(2)},'' 2007.

\bibitem{Bourget:2015upj}
A.~Bourget and J.~Troost, ``{On the $ \mathcal{N}={1}^{\ast } $ gauge theory on
  a circle and elliptic integrable systems},'' {\em JHEP}, vol.~01, p.~097,
  2016.

\bibitem{Ramond:1996ku}
P.~Ramond, ``{Superalgebras in N=1 gauge theories},'' {\em Phys. Lett.},
  vol.~B390, pp.~179--184, 1997.

\bibitem{Distler:1996ub}
J.~Distler and A.~Karch, ``{N=1 dualities for exceptional gauge groups and
  quantum global symmetries},'' {\em Fortsch. Phys.}, vol.~45, pp.~517--533,
  1997.

\bibitem{Karch:1997jp}
A.~Karch, ``{More on N=1 selfdualities and exceptional gauge groups},'' {\em
  Phys. Lett.}, vol.~B405, pp.~280--286, 1997.

\bibitem{Cho:1997am}
P.~L. Cho, ``{Moduli in exceptional SUSY gauge theories},'' {\em Phys. Rev.},
  vol.~D57, pp.~5214--5223, 1998.

\bibitem{Pouliot:2001iw}
P.~Pouliot, ``{Spectroscopy of gauge theories based on exceptional Lie
  groups},'' {\em J. Phys.}, vol.~A34, pp.~8631--8658, 2001.

\bibitem{Csaki:1997aw}
C.~Csaki and H.~Murayama, ``{Discrete anomaly matching},'' {\em Nucl. Phys.},
  vol.~B515, pp.~114--162, 1998.

\bibitem{Danielsson:1995zi}
U.~H. Danielsson and B.~Sundborg, ``{Exceptional equivalences in N=2
  supersymmetric Yang-Mills theory},'' {\em Phys. Lett.}, vol.~B370,
  pp.~83--94, 1996.

\bibitem{Abolhasani:1996np}
M.~R. Abolhasani, M.~Alishahiha, and A.~M. Ghezelbash, ``{The Moduli space and
  monodromies of the N=2 supersymmetric Yang-Mills theory with any Lie gauge
  groups},'' {\em Nucl. Phys.}, vol.~B480, pp.~279--295, 1996.

\bibitem{Aharony:2013dha}
O.~Aharony, S.~S. Razamat, N.~Seiberg, and B.~Willett, ``{3d dualities from 4d
  dualities},'' {\em JHEP}, vol.~07, p.~149, 2013.

\bibitem{Aharony:2013kma}
O.~Aharony, S.~S. Razamat, N.~Seiberg, and B.~Willett, ``{3$d$ dualities from
  4$d$ dualities for orthogonal groups},'' {\em JHEP}, vol.~08, p.~099, 2013.

\bibitem{Pouliot:1995zc}
P.~Pouliot, ``{Chiral duals of nonchiral SUSY gauge theories},'' {\em Phys.
  Lett.}, vol.~B359, pp.~108--113, 1995.

\bibitem{Witten:1982df}
E.~Witten, ``{Constraints on Supersymmetry Breaking},'' {\em Nucl. Phys.},
  vol.~B202, p.~253, 1982.

\bibitem{Morozov:1987hy}
A.~{\relax Yu}. Morozov, M.~A. Olshanetsky, and M.~A. Shifman, ``{Gluino
  Condensate in Supersymmetric Gluodynamics},'' {\em Sov. Phys. JETP}, vol.~67,
  p.~222, 1988.
\newblock [Zh. Eksp. Teor. Fiz.94,18(1988)].

\bibitem{Witten:1997bs}
E.~Witten, ``{Toroidal compactification without vector structure},'' {\em
  JHEP}, vol.~02, p.~006, 1998.

\bibitem{Weinberg:2006rq}
E.~J. Weinberg and P.~Yi, ``{Magnetic Monopole Dynamics, Supersymmetry, and
  Duality},'' {\em Phys. Rept.}, vol.~438, pp.~65--236, 2007.

\bibitem{Shnir:2015hha}
{\relax Ya}.~Shnir and G.~Zhilin, ``{G2 monopoles},'' {\em Phys. Rev.},
  vol.~D92, p.~045025, 2015.

\bibitem{Matsudo:2016cui}
R.~Matsudo and K.-I. Kondo, ``{Gauge-covariant decomposition and magnetic
  monopole for G(2) Yang-Mills field},'' {\em Phys. Rev.}, vol.~D94, no.~4,
  p.~045004, 2016.

\bibitem{Lee:1997vp}
K.-M. Lee and P.~Yi, ``{Monopoles and instantons on partially compactified
  D-branes},'' {\em Phys. Rev.}, vol.~D56, pp.~3711--3717, 1997.

\bibitem{Lee:1998vu}
K.-M. Lee, ``{Instantons and magnetic monopoles on R**3 x S**1 with arbitrary
  simple gauge groups},'' {\em Phys. Lett.}, vol.~B426, pp.~323--328, 1998.

\bibitem{Kapustin:2005py}
A.~Kapustin, ``{Wilson-'t Hooft operators in four-dimensional gauge theories
  and S-duality},'' {\em Phys. Rev.}, vol.~D74, p.~025005, 2006.

\bibitem{Callias:1977kg}
C.~Callias, ``{Index Theorems on Open Spaces},'' {\em Commun. Math. Phys.},
  vol.~62, pp.~213--234, 1978.

\bibitem{Weinberg:1979zt}
E.~J. Weinberg, ``{Fundamental Monopoles and Multi-Monopole Solutions for
  Arbitrary Simple Gauge Groups},'' {\em Nucl. Phys.}, vol.~B167, pp.~500--524,
  1980.

\bibitem{deBoer:1997kr}
J.~de~Boer, K.~Hori, and Y.~Oz, ``{Dynamics of N=2 supersymmetric gauge
  theories in three-dimensions},'' {\em Nucl. Phys.}, vol.~B500, pp.~163--191,
  1997.

\bibitem{Affleck:1982as}
I.~Affleck, J.~A. Harvey, and E.~Witten, ``{Instantons and (Super)Symmetry
  Breaking in (2+1)-Dimensions},'' {\em Nucl. Phys.}, vol.~B206, pp.~413--439,
  1982.

\bibitem{Intriligator:2013lca}
K.~Intriligator and N.~Seiberg, ``{Aspects of 3d N=2 Chern-Simons-Matter
  Theories},'' {\em JHEP}, vol.~07, p.~079, 2013.

\bibitem{Aharony:1997bx}
O.~Aharony, A.~Hanany, K.~A. Intriligator, N.~Seiberg, and M.~J. Strassler,
  ``{Aspects of N=2 supersymmetric gauge theories in three-dimensions},'' {\em
  Nucl. Phys.}, vol.~B499, pp.~67--99, 1997.

\bibitem{Nahm:1983sv}
W.~Nahm, ``{SELFDUAL MONOPOLES AND CALORONS},'' in {\em {Group Theoretical
  Methods in Physics: PROCEEDINGS, 12TH INTERNATIONAL COLLOQUIUM, TRIESTE,
  ITALY, SEPTEMBER 5-11, 1983}}, 1983.

\bibitem{Bhattacharya:2008bja}
J.~Bhattacharya and S.~Minwalla, ``{Superconformal Indices for N = 6 Chern
  Simons Theories},'' {\em JHEP}, vol.~01, p.~014, 2009.

\bibitem{Kim:2009wb}
S.~Kim, ``{The Complete superconformal index for N=6 Chern-Simons theory},''
  {\em Nucl. Phys.}, vol.~B821, pp.~241--284, 2009.
\newblock [Erratum: Nucl. Phys.B864,884(2012)].

\bibitem{Imamura:2011su}
Y.~Imamura and S.~Yokoyama, ``{Index for three dimensional superconformal field
  theories with general R-charge assignments},'' {\em JHEP}, vol.~04, p.~007,
  2011.

\bibitem{Imamura:2011uj}
Y.~Imamura, D.~Yokoyama, and S.~Yokoyama, ``{Superconformal index for large N
  quiver Chern-Simons theories},'' {\em JHEP}, vol.~08, p.~011, 2011.

\bibitem{Kapustin:2011jm}
A.~Kapustin and B.~Willett, ``{Generalized Superconformal Index for Three
  Dimensional Field Theories},'' 2011.

\bibitem{Bashkirov:2011vy}
D.~Bashkirov, ``{Aharony duality and monopole operators in three dimensions},''
  2011.

\bibitem{Kapustin:2011vz}
A.~Kapustin, H.~Kim, and J.~Park, ``{Dualities for 3d Theories with Tensor
  Matter},'' {\em JHEP}, vol.~12, p.~087, 2011.

\bibitem{Kim:2013cma}
H.~Kim and J.~Park, ``{Aharony Dualities for 3d Theories with Adjoint
  Matter},'' {\em JHEP}, vol.~06, p.~106, 2013.

\bibitem{Spiridonov:2009za}
V.~P. Spiridonov and G.~S. Vartanov, ``{Elliptic Hypergeometry of
  Supersymmetric Dualities},'' {\em Commun. Math. Phys.}, vol.~304,
  pp.~797--874, 2011.

\bibitem{Kapustin:2009kz}
A.~Kapustin, B.~Willett, and I.~Yaakov, ``{Exact Results for Wilson Loops in
  Superconformal Chern-Simons Theories with Matter},'' {\em JHEP}, vol.~03,
  p.~089, 2010.

\bibitem{Goddard:1976qe}
P.~Goddard, J.~Nuyts, and D.~I. Olive, ``{Gauge Theories and Magnetic
  Charge},'' {\em Nucl. Phys.}, vol.~B125, pp.~1--28, 1977.

\bibitem{Cremonesi:2015dja}
S.~Cremonesi, ``{The Hilbert series of 3d ${\boldsymbol{\mathcal{N}}}=2$
  Yang-Mills theories with vectorlike matter},'' {\em J. Phys.}, vol.~A48,
  no.~45, p.~455401, 2015.

\bibitem{Hanany:2015via}
A.~Hanany, C.~Hwang, H.~Kim, J.~Park, and R.-K. Seong, ``{Hilbert Series for
  Theories with Aharony Duals},'' {\em JHEP}, vol.~11, p.~132, 2015.
\newblock [Addendum: JHEP04,064(2016)].

\bibitem{Benvenuti:2006qr}
S.~Benvenuti, B.~Feng, A.~Hanany, and Y.-H. He, ``{Counting BPS Operators in
  Gauge Theories: Quivers, Syzygies and Plethystics},'' {\em JHEP}, vol.~11,
  p.~050, 2007.

\bibitem{Feng:2007ur}
B.~Feng, A.~Hanany, and Y.-H. He, ``{Counting gauge invariants: The Plethystic
  program},'' {\em JHEP}, vol.~03, p.~090, 2007.

\bibitem{Pouliot:1998yv}
P.~Pouliot, ``{Molien function for duality},'' {\em JHEP}, vol.~01, p.~021,
  1999.

\bibitem{Csaki:2014cwa}
C.~Cs\'{a}ki, M.~Martone, Y.~Shirman, P.~Tanedo, and J.~Terning, ``{Dynamics of
  3D SUSY Gauge Theories with Antisymmetric Matter},'' {\em JHEP}, vol.~08,
  p.~141, 2014.

\bibitem{Amariti:2015kha}
A.~Amariti, C.~Cs\'{a}ki, M.~Martone, and N.~R.-L. Lorier, ``{From 4D to 3D
  chiral theories: Dressing the monopoles},'' {\em Phys. Rev.}, vol.~D93,
  no.~10, p.~105027, 2016.

\bibitem{Behrends:1962zz}
R.~E. Behrends, J.~Dreitlein, C.~Fronsdal, and W.~Lee, ``{Simple Groups and
  Strong Interaction Symmetries},'' {\em Rev. Mod. Phys.}, vol.~34, pp.~1--40,
  1962.

\bibitem{Slansky:1981yr}
R.~Slansky, ``{Group Theory for Unified Model Building},'' {\em Phys. Rept.},
  vol.~79, pp.~1--128, 1981.

\bibitem{Ruben}
R.~Arenas, ``{Constructing a Matrix Representation of the Lie Group G2},''
  Master's thesis, Department of Mathematics, Harvey Mudd college.

\end{thebibliography}

\end{document}